\documentclass[12pt]{article}
\usepackage[utf8]{inputenc}

\usepackage[left=3cm, right=3cm, top=2.5cm, bottom=2.5cm, bindingoffset=0cm]{geometry}

\usepackage{tikz, tikz-cd}
\usetikzlibrary{calc,intersections,through,backgrounds}
\usepackage{amsmath, amssymb}
\usepackage{amsthm, mathtools, hyperref}

\mathtoolsset{showonlyrefs}

\newtheorem{lemma}{Lemma}[section]

\newtheorem{theorem}[lemma]{Theorem}
\newtheorem{corollary}[lemma]{Corollary}

\theoremstyle{definition}
\newtheorem{remark}[lemma]{Remark}

\newtheorem{example}[lemma]{Example}

\usetikzlibrary{arrows.meta}

\usetikzlibrary{positioning}
\usetikzlibrary{decorations.text}
\usetikzlibrary{decorations.pathmorphing}
\usetikzlibrary{decorations.markings}

\usepackage{tkz-euclide}

\newcounter{ai}

\newcounter{mg}

\newcommand{\eps}{{\varepsilon}}

\newcommand{\C}{{\mathbf C}}

\newcommand{\R}{{\mathbf R}}
\newcommand{\Z}{{\mathbf Z}}

\def\can{{\mathcal N}}

\def\EE{{\mathcal E}}
\def\FF{{\mathcal F}}

\newcommand{\Q}{\mathcal{Q}}
\newcommand{\K}{\mathbf{K}}
\def\corank{\operatorname{corank}}

\usepackage[numbers, sort&compress]{natbib}

  \tikzset{->-/.style={decoration={
  markings,
  mark=at position .5 with {\arrow{>}}},postaction={decorate}}}
    \tikzset{-<-/.style={decoration={
  markings,
  mark=at position .5 with {\arrow{<}}},postaction={decorate}}}

    \tikzset{-->-/.style={decoration={
  markings,
  mark=at position .8 with {\arrow{>}}},postaction={decorate}}}

\title{Integrable systems and cluster algebras}
\author{Michael Gekhtman\thanks{Department of Mathematics, University of Notre Dame, e-mail: \tt{mgekhtma@nd.edu}}\,\, and Anton Izosimov\thanks{Department of Mathematics, University of Arizona, e-mail: \tt{izosimov@math.arizona.edu}}}

\date{}

\begin{document}

\maketitle

\begin{abstract}
    We review several constructions of integrable systems with an underlying cluster algebra structure, in particular the Gekhtman-Shapiro-Tabachnikov-Vainshtein construction based on perfect networks and the Goncharov-Kenyon approach based on the dimer model. We also discuss results of Galashin and Pylyavskyy on integrability of $T$-systems.
\end{abstract}

\tableofcontents

\section{Introduction}
Cluster algebras  are axiomatically defined commutative rings
equipped with a distinguished set of generators (cluster
variables)
subdivided into overlapping subsets (clusters) of the same
cardinality subject to certain polynomial relations. 
They were discovered by Fomin and Zelevinsky \cite{FZ} just over twenty years ago and quickly found applications in various fields. There is, by now, a vast literature devoted to the theory and applications of cluster algebras, including several books, both published \cite{GSV3,Marsh, Tomoki} and evolving on {\tt arXiv} \cite{FWZ}.
Relations of cluster type have been
observed in many areas of mathematics (Gauss' {\em Pentagramma Mirificum}, Pl\"ucker  and Ptolemy
relations, Stokes and wall-crossing  phenomena, Somos sequences  and KP integrable hierarchy 
to name just a few examples). Rapid development of the cluster 
algebra theory
revealed 
connections to Grassmannians, 
quiver representations, canonical bases,
(higher) Teichm\"uller theory,
Poisson geometry, 3D gauge theories 
and many other branches of
  mathematics and mathematical physics.
In the latter direction, several advances were made in the study of classical and quantum continuous and discrete integrable systems arising in the context of cluster algebras. Some of these  systems are defined on Poisson submanifolds of Poisson–Lie groups and their Poisson homogeneous spaces that support a cluster structure \cite{Backlund, FM, Williams}. One such example is Toda flows on double Bruhat cells in simple Lie groups  \cite{Backlund}. More generally, one considers discrete maps obtained as  distinguished sequences of cluster maps associated with quivers of special kinds. This phenomenon was first observed in \cite{fordy2011cluster}, where integrability properties of Somos recurrences were studied in the context of cluster transformations. Some of the recurrences arising in the study of representations of quantized affine algebras, e.g. $Q$-, $T$- and $Y$-systems \cite{DiFraKed, Kedem, Inoue}, can also be realized this way. In addition, many natural maps that can be described via elementary projective geometry have found cluster algebraic interpretations, the most prominent example being R.Schwartz's {\em pentagram map} and its various generalizations \cite{glick2011pentagram, Sch, GSTV, glick2016meshes}. All of the examples just mentioned are interconnected, and it became common to refer to them as {\em cluster integrable systems}, although one would be hard pressed to give a precise general definition of what a cluster integrable system is. We will not attempt to come up with such a definition here. Neither will we be able to comprehensively review the vast literature on interactions between the theories of cluster algebras and integrable systems. Instead, we chose to describe in some detail two constructions that generate large families of integrable systems. One, outlined in Section \ref{xysect}, is due to Gekhtman, Shapiro, Tabachnikov  and Vainshtein \cite{GSTV} and is centered on weighted directed networks on surfaces. The other, due to Goncharov and Kenyon \cite{GK}, is based on the dimer model. It is described in Section~\ref{sec:GK}. The equivalence of the approaches above was recently established in \cite{izosimov2022dimers}. Sections~\ref{xysect} and \ref{sec:GK} are preceded by background information on Poisson manifolds and integrable systems (Section~\ref{sec:pmis}) and quiver mutations and cluster maps (Section~\ref{sec:qmrs}). In Section~\ref{sec:GP}, we discuss the series of works by Galashin and Pylyavskyy \cite{GalPylAJM,GalPylDM,GalPylMZ} on classification of integrability patterns of $T$-systems associated with bipartite recurrent quivers in terms of quivers' combinatorial properties. The concluding Section \ref{sec:further} contains a very incomplete list of examples of cluster integrable systems arising from the constructions discussed, as well as further results extending these constructions.

\paragraph{Acknowledgements.} A.I. was supported by NSF grant DMS-2008021. M.G. was supported by NSF grant DMS-2100785. The authors are grateful to P.\,Pylyavskyy for fruitful discussions.

\section{Poisson manifolds and integrable systems}\label{sec:pmis}


In this section we define Poisson manifolds and integrable systems. All definitions are given in the real setting. The general definition of a holomorphic Poisson manifold is more involved due to possible absence of global holomorphic functions. However, everything works without significant modifications when the manifold in question is an open subset of $\C^n$. We refer the reader to the book \cite{Vanh} for a detailed discussion of the concepts presented below.

\paragraph{Poisson manifolds, maps, and submanifolds. }A smooth manifold $M$ is called a \textit{Poisson manifold} if its algebra $ C^\infty(M)$ of smooth functions is endowed with a Lie bracket $\{\, ,\}$, called the \emph{Poisson bracket}, which is a derivation in first (and hence in each) argument:
$$
\{fg,h\} = f\{g,h\} + \{f,h\}g
$$
for all $f,g,h \in C^\infty(M)$.
Such a structure is determined by a smooth bivector $\pi$ on $M$, called the \emph{Poisson bivector}, such that
$$
\{f,g\} = \pi(df, dg)
$$
for any $f,g \in C^\infty(M)$. In local coordinates, the components of $\pi$ are just pairwise Poisson brackets of coordinate functions. In particular, if the manifold $M$ is an open subset of $\R^n$, then a Poisson bracket on $M$ is determined by pairwise Poisson brackets of linear functions.

A smooth map $f \colon M \to N$ between Poisson manifolds is called a \emph{Poisson map} if the induced map $f^* \colon C^\infty(N) \to C^\infty(M)$ takes a Poisson bracket on $N$ to the Poisson bracket on $M$. A submanifold $N \subset M$ is a \textit{Poisson submanifold} if there exists a Poisson structure on $N$ (which is then necessarily unique and is called the \emph{induced Poisson structure} on $N$) such that the inclusion map $i \colon N \to M$ is Poisson.

\paragraph{The rank of a Poisson structure, symplectic leaves, and Casimirs.}
 The \emph{rank of a Poisson bracket at a point} $x \in M$ is the rank of the corresponding Poisson bivector considered as a bilinear function on $T^*_xM$. If the rank  at every point coincides with the dimension of $M$, then the Poisson tensor $\pi$ is invertible, and its inverse $\pi^{-1}$ is a \emph{symplectic form}, i.e. a closed non-degenerate $2$-form on $M$. Conversely, any symplectic manifold can be viewed as a Poisson manifold whose Poisson structure has maximal rank. In general, when the rank of a Poisson structure is allowed to drop, a Poisson manifold can be represented as a disjoint union of \emph{symplectic leaves}, i.e. Poisson submanifolds whose induced Poisson structure is invertible and hence determined by a symplectic structure. The dimension of a symplectic leaf through every point coincides with the rank of the Poisson tensor at that point. Under sufficiently mild assumptions, e.g. if the manifold and the Poisson tensor are real-analytic, the rank is the same at almost every point, so almost all symplectic leaves have the same codimension $k$ equal to the corank of the Poisson structure at a generic point. In that situation, at least locally, symplectic leaves can be described as joint level sets of $k$ functionally independent (i.e. having almost everywhere linearly independent differentials) \emph{Casimir functions} i.e. functions $f \in C^\infty(M)$ such that $\{f,g\} = 0$ for all $g \in C^\infty(M)$. All Poisson structures $\pi$ considered in this paper admit $k = \corank \pi$ globally defined Casimir functions whose joint level sets are the symplectic leaves.

 \paragraph{Continuous integrable systems.} Two functions $f,g \in C^\infty(M)$ \emph{Poisson commute}, or are \emph{in involution}, if $\{f,g\} = 0$ everywhere on $M$. If $M$ has dimension $n$, and the corank of the Poisson structure at generic points is equal to $k$, then the maximal number of Poisson-commuting functionally-independent functions is $m = \frac{1}{2}(n+k)$. Such a maximal collection $f_1, \dots, f_m$ is called a (continuous-time) \emph{integrable system} (while functions $f_1, \dots, f_m$ themselves will be referred to below as \emph{Hamiltonians}). Typically, $f_1, \dots, f_k$ are Casimirs, while the remaining $\frac{1}{2}(n-k)$ functions $f_{k+1}, \dots, f_m$ are independent on generic symplectic leaves. Each of those functions $f_{k+1}, \dots, f_m$ gives rise to a Hamiltonian vector field $v_i$ on $M$ defined by $v_i := \pi df_i$ (while Hamiltonian vector fields associated to Casimirs are zero). The fields $v_i$ pairwise commute and are tangent to the foliation defined by the functions  $f_1, \dots, f_m$. Under certain additional assumptions, one shows that the leaves of that foliation are tori, on which the fields $v_i$ become linear (Liouville-Arnold theorem). That allows to explicitly integrate the fields $v_i$ by means of quadratures (which in most examples boil down to special functions such as theta functions), hence the name: integrable systems.

 \paragraph{Discrete integrable systems.} 
 A \emph{discrete integrable system} (or an \emph{integrable map}) is a Poisson automorphism $\phi \colon M \to M$ which has a maximal collection of Poisson-commuting invariant functions (first integrals) $f_1, \dots, f_m \in C^\infty(M)$. In particular, if $\phi$ is an integrable map, then its first integrals form a continuous integrable system. Conversely, one can start with a continuous integrable system $f_1, \dots, f_m$ and take the time-$1$ shift along the trajectories of $v_i := \pi df_i$ as a discrete system. However, this usually does not produce interesting discrete systems, in particular because time-$1$ shifts are hard to write down explicitly. Discrete systems that we will discuss in this paper are, on the contrary, explicitly given birational maps (and, in fact, \emph{cluster transformations}).






\section{Quivers, mutations, and cluster maps}\label{sec:qmrs}


In this section, without explicitly defining cluster algebras, we review the basic notions of the cluster theory that will be needed in what follows. 
In-depth expositions of the theory of cluster algebras can be found in \cite{GSV3,Marsh,FWZ}.

\paragraph{Quivers and quiver mutations.} A \textit{quiver} is a directed graph. In what follows, all quivers will be finite, with vertices labeled by integers $1, \dots, n$, and without loops or directed cycles of length $2$.

Given a quiver $\Q$, its \textit{mutation} at vertex $i$ is the following modification of~$\Q$:
\begin{enumerate}
\item For every directed path $j \to i \to k$ of length $2$, add an arrow from $j$ to $k$.
\item Reverse all arrows adjacent to the vertex $i$.
\item Remove newly formed oriented cycles of length $2$, one at a time until there are no such cycles left.
\end{enumerate}
\begin{figure}[t]
\centering
\begin{tikzpicture}[]
\node () at (0,0)
{
\begin{tikzpicture}[]
\node () at (-1,1) {$\Q$};
\node  (A) at (0,0) {$1$};
\node (B) at (2,0) {$2$};
\node (C) at (0,1) {$3$};
\node(D) at (-2,0) {$4$};
\draw [->-,  double distance=3pt] (A) to (B);
\draw [->] (B) -- (C);
\draw [->] (C) -- (A);
\draw [->] (A) -- (D);
\end{tikzpicture}
};
\node () at (3,0) {$\longrightarrow$};
\node () at (6,0)
{
\begin{tikzpicture}[]
\node () at (2,1) {$\Q'$};
\node  (A) at (0,0) {$1$};
\node (B) at (2,0) {$2$};
\node (C) at (0,1) {$3$};
\node(D) at (-2,0) {$4$};
\draw [-<-, double distance=3pt] (A) to (B);
\draw [->-] (C) -- (B);
\draw [<-] (C) -- (A);
\draw [->] (C) -- (D);
\draw [->] (D) -- (A);
\end{tikzpicture}
};
\end{tikzpicture}
\caption{Quiver mutation at vertex $1$.}\label{FigQM}
\end{figure}
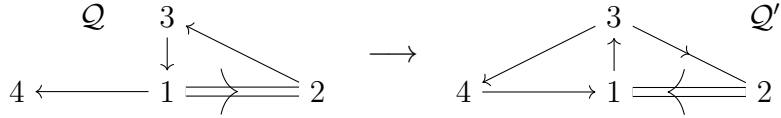
The result of such a mutation is a new quiver $\Q'$ with the same vertex set as $\Q$. An example of a quiver mutation is shown in Figure \ref{FigQM}. Note that a quiver mutation is an involution: if we mutate a quiver at the same vertex twice, we get the initial quiver. Each quiver mutation gives rise to two birational involutions: an \emph{$X$-mutation} and a \emph{$Y$-mutation}, both described below. 
\paragraph{$X$-type cluster mutations.} Given a quiver $\Q$, let $X_\Q$ be the space of functions from its vertex set $\{1, \dots, n\}$ to the multiplicative group $\K^*$ of the ground field~$\K$ (where $\K$ = $\R$ or $\C$). The set $X_\Q$ is an algebraic torus $(\K^*)^{n}$ with coordinates $x_1, \dots, x_n$ given by evaluation of functions at vertices of $\Q$: for $\xi \in X_{\Q}$, one defines $x_i(\xi) := \xi(i)$. Since the variables $x_i$ are indexed by vertices of $\Q$, in what follows we often identify vertices with the corresponding variables.

Assume that a quiver $\Q'$ is obtained from  $\Q$ by means of mutation at vertex~$i$. The corresponding \textit{$X$-mutation} is a birational involution $\mu^X_i \colon X_{\Q } \to X_{\Q'} $, 
$
(x_1, \dots, x_n) \mapsto (x_1', \dots, x_n') ,
$ with $x_j' = x_j$ for $i \neq j$, and $x_i'$ being defined by the following \textit{exchange relation}:
$$
x_i x_i' = \prod_{j =1}^n x_j^{\#\{j \to i\}} \,+\,  \prod_{j =1}^n x_j^{\#\{i \to j\}},
$$
where $\#\{j \to i\}$ stands for the number of arrows in $\Q$ from vertex $j$ to vertex~$i$.
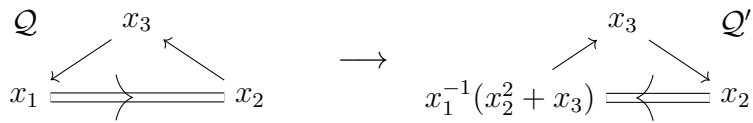
\begin{figure}[b]
\centering
\begin{tikzpicture}[]
\node () at (0,0)
{
\begin{tikzpicture}[]
\node () at (0,1) {$\Q$};
\node  (A) at (0,0) {$x_1$};
\node (B) at (3,0) {$x_2$};
\node (C) at (1.5,1) {$x_3$};
\draw [->-, double distance=3pt ] (A) to (B);
\draw [->] (B) -- (C);
\draw [->] (C) -- (A);
\end{tikzpicture}
};
\node () at (3,0) {$\longrightarrow$};
\node () at (6,0)
{
\begin{tikzpicture}[]
\node () at (3,1) {$\Q'$};
\node  (A) at (0,0) {$x_1^{-1}(x_2^2 + x_3)$};
\node (B) at (3,0) {$x_2$};
\node (C) at (1.5,1) {$x_3$};
\draw [-<-, double distance=3pt] (A) to (B);
\draw [->] (C) -- (B);
\draw [<-] (C) -- (A);
\end{tikzpicture}
};
\end{tikzpicture}
\caption{$X$-type mutation at vertex $x_1$.}\label{FigX}
\end{figure}
We depict $X$-mutations as shown in Figure \ref{FigX}. The labels at vertices of the initial quiver $\Q$ are the corresponding $X$-variables $x_j$ while the labels at vertices of the mutated quiver $\Q'$ are pull-backs $(\mu_i^X)^*x_j'$.

Now assume we have sequence of quivers $\Q  \to \dots \to \tilde \Q$ where each quiver is obtained from the previous one by mutation. Suppose also that we have an isomorphism of directed graphs $\psi \colon \tilde \Q \to \Q$. Then the composition $X_{\Q} \to \dots \to X_{\tilde \Q}$ of $X$-mutations, followed by the map $X_{\tilde \Q} \to X_{\Q}$ induced by the isomorphism $\psi$, is a birational map of $X_{\Q}$ onto itself. We call such a map an $X$-type \textit{cluster map}, cf. \cite{fordy2011cluster, nakanishi2011periodicities}. Put differently, a cluster map is such a sequence of mutations which, after permutation of vertices, restores the initial quiver. \begin{example} Switching two bottom-most vertices of the mutated quiver $\Q'$ in Figure \ref{FigX}, we obtain the initial quiver $\Q$. Hence, we get a cluster map $X_\Q \to X_\Q$ given by
$
(x_1, x_2, x_3) \mapsto (x_2, x_1^{-1}(x_2^2 + x_3), x_3).
$
\end{example}

\paragraph{Canonical presymplectic structure.}

We note in passing that the tori $X_\Q$ defined above carry a natural presymplectic structure (i.e. a closed $2$-form). It is given by
\begin{equation}\label{eq:presym}
\omega_\Q := \sum_{i,j =1}^n b_{ij}\frac{dx_i \wedge dx_j}{x_ix_j},
 \end{equation}
 where $b_{ij}  := \#\{i \to j \} -  \#\{j \to i \}$ is the signed adjacency matrix of $\Q$ (the so-called \emph{exchange matrix}).  
The pre-symplectic structures $\omega_\Q$ behave in a covariant way under $X$-mutations \cite[Chapter 6]{GSV3}. Namely, for a mutation $\mu^X_i \colon X_\Q \to X_{\Q'}$ we have
$
(\mu^X_i)^* \omega_{\Q'} = \omega_\Q.
$
Therefore, any $X$-type cluster transformation $\phi^X \colon X_\Q \to X_\Q$ preserves the pre-symplectic structure $\omega_\Q$.

\paragraph{$Y$-type cluster mutations.}
We now define another algebraic torus associated with a quiver $\Q$, denoted by $Y_\Q$. As a set, $Y_\Q$ is the same as $X_\Q$, i.e. the space of functions $\{1, \dots, n\} \to \K^*$. Coordinates in $Y_\Q$ are again given by evaluation at vertices and are denoted by $y_i$.

Assume that a quiver $\Q'$ is obtained from  $\Q$ by means of mutation at vertex~$i$.
A \textit{$Y$-mutation} is a birational involution $\mu^Y_i \colon Y_{\Q } \to Y_{\Q'} $ given by
\begin{equation*}
y_j' = \left[\begin{aligned} &y_j^{-1}, \quad \mbox{if $j = i$},\\
&y_j(1 + y_i^{-1})^{-k}, \quad \mbox{if there are $k$ arrows from vertex $i$ to vertex $j$},\\
&y_j(1 + y_i)^k, \quad \mbox{if there are $k$ arrows from vertex $j$ to vertex $i$},\\
&y_j,\quad \mbox{in all other cases}.
\end{aligned} \right.
\end{equation*}
The map above is a particular case of the {\em coefficient mutation} introduced in \cite{FZ} as part of the general definition of a cluster algebra.

The relation between $X$ and $Y$ mutations is as follows. Consider the map $p_\Q \colon X_\Q \to Y_\Q$ (the \textit{ensemble map}) given by
\begin{equation}\label{eq:xymap}
y_i = \prod_{j=1}^n x_j^{b_{ji}}
\end{equation}
where $b_{ij}$ is the signed adjacency matrix of $\Q$
\footnote{The term "ensemble map" was coined in \cite{FG}, but the map itself was introduced earlier in two different contexts in \cite{GSV} and \cite{FZ4}}.
Then, for the quiver $\Q'$ obtained from  $\Q$ by means of mutation at vertex~$i$, the following diagram commutes:
\begin{equation}\label{eq:enscd}
\begin{tikzcd}
X_\Q \arrow[r, "\mu_i^X"] \arrow[d, "p_\Q"]
& X_{\Q'} \arrow[d, "p_{\Q'}"] \\
Y_\Q \arrow[r, "\mu_i^Y" ]
&  Y_{\Q'}.
\end{tikzcd}
\end{equation}
In particular, if the signed adjacency matrix of $\Q$ is invertible (which is a property invariant under mutations), then so is the ensemble map $p_\Q$, which allows one to identify $X$ and $Y$ mutations.

$Y$-type cluster maps are defined analogously to $X$-type: they are birational maps $Y_\Q \to Y_\Q$ which can be represented as sequences of $Y$-mutations followed by a permutation of vertices. If $\phi^X$ and $\phi^Y$ are $X$ and $Y$ cluster maps given by the same sequences of mutations and same permutation, then from~\eqref{eq:enscd} we get 
$
\phi^Y \circ p_\Q = p_\Q \circ \phi^X.
$
In particular, if the signed adjacency matrix of $\Q$ is invertible, then $Y$-type cluster maps are conjugate to $X$-type cluster maps.
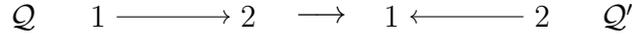
\begin{figure}[t]
\centering
\begin{tikzpicture}[]
\node () at (0,0)
{
\begin{tikzpicture}[]
\node () at (-1,0) {$\Q$};
\node  (A) at (0,0) {$1$};
\node (B) at (2,0) {$2$};
\draw [->] (A) to (B);
\end{tikzpicture}
};
\node () at (2.5,0) {$\longrightarrow$};
\node () at (5,0)
{
\begin{tikzpicture}[]
\node () at (3,0) {$\Q'$};
\node  (A) at (0,0) {$1$};
\node (B) at (2,0) {$2$};
\draw [<-] (A) to (B);
\end{tikzpicture}
};
\end{tikzpicture}
\caption{Quiver mutation at vertex $1$.}\label{FigQM2}
\end{figure}
\begin{example}
Consider the quiver mutation in Figure \ref{FigQM2}. The quivers $\Q$ and $\Q'$ are isomorphic, giving rise to an $X$-type cluster map
$
(x_1, x_2) \to (x_2, x_1^{-1}(1+x_2))
$
and $Y$-type cluster map
$
(y_1, y_2) \to (y_2(1+y_1^{-1})^{-1},y_1^{-1}).
$
These two maps are related by a change of variables given by the ensemble map~\eqref{eq:xymap}:
$y_1 = x_2^{-1}, y_2 = x_1.$
\end{example}
\paragraph{Poisson property of $Y$-mutations.} The tori $Y_\Q$ carry a canonical Poisson structure given by
\begin{equation}\label{eq:pb}
\{y_i, y_j\}_\Q = b_{ij}y_iy_j,
\end{equation}
where $b_{ij}$ is the signed adjacency matrix of $\Q$. These Poisson structures behave nicely under $Y$-mutations in the sense that any such mutation $\mu_i^Y$ is a (birational) Poisson map $(Y_\Q, \{\, ,\}_\Q) \to (Y_{\Q'}, \{\, ,\}_{\Q'})$. In particular, $Y$-type cluster maps are birational Poisson automorphisms.

The relation between the Poisson bracket \eqref{eq:pb} and pre-symplectic structure~\eqref{eq:presym} is as follows. The image of the ensemble map $p_\Q \colon X_\Q \to Y_\Q$ defined by \eqref{eq:xymap} is a symplectic leaf of the bracket \eqref{eq:pb}. Furthermore, the pull-back of the  symplectic structure on that leaf by $p_Q$ coincides with the form~\eqref{eq:presym}. In particular, when the signed adjacency matrix of the quiver $\Q$ is invertible, pre-symplectic structure~\eqref{eq:presym} is symplectic and the Poisson bracket \eqref{eq:pb} is just its inverse, rewritten in $Y$-coordinates.

We will also occasionally need brackets of the form \eqref{eq:pb} with half-integer numbers $b_{ij}$. Such brackets can be modeled using quivers with \textit{half-edges}. A half-edge is an edge to which we assign a weight $1/2$. A half-edge from vertex $i$ to vertex $j$ gives a contribution $+1/2$ to $b_{ij}$ and $-1/2$ to $b_{ji}$.

\section{Cluster integrable systems from weighted directed networks} \label{xysect}
In this section we describe the construction of cluster integrable systems based on weighted directed networks \cite{GSTV}.
\subsection{Continuous systems}\label{xyint}
\paragraph{Perfect networks on a cylinder.}
We begin with a brief overview of the theory of weighted directed networks on surfaces with boundary,  see \cite{Po,GSV3} for details. We will concentrate on networks on a cylinder (equivalently, annulus) $\mathcal{C}$ that we position horizontally with one boundary circle on the left and another on the right.

 Let $\can$ be a directed graph with vertex set $V$ and edge set $E$ embedded in $\mathcal{C}$ so that connected components of $\mathcal C \setminus \can$ are contractible.
Suppose that $\can$ has $2n$
{\it boundary vertices\/}, each of degree one: $n$ {\it sources\/} on the left boundary circle and $n$ {\it sinks\/} on the right
boundary circle.
All the internal vertices of $\can$ have degree~$3$ and are of two types: either they have exactly one
incoming edge (\emph{white vertices}), or exactly one outgoing edge (\emph{black vertices})\footnote{There is a also a more general version of this definition, where vertices do not have to be $3$-valent, but are still required to have either exactly one incoming edge (white vertices) or exactly one outgoing edge (black vertices).}.
To each edge $e\in E$ we assign an {\it edge weight\/} $w_{e}\in \K^*$.
Such a weighted graph $\can$ on a cylinder is called a {\it perfect network\/}. The points of the {\it space of edge weights\/} $\EE_\can$
can be considered as copies of $\can$  with edges weighted by elements of~$\K^*$.

\paragraph{The boundary measurement map.} A \emph{cut} $\rho$ of a perfect network $\can$ is an oriented curve without self-intersections that joins  the left
and the right boundary circles and does not contain vertices of $\can$. From now on we will assume that the choice of the cut is fixed. Assign an independent variable $\lambda$ to the cut $\rho$. 
 The weight of a directed path $P$ between a source and a sink
is defined as a signed product of the weights of all edges along the path times  $\lambda^d$, where $d$
is the intersection index of $\rho$ and $P$, and the sign is basically the rotation number of the path mod $2$. 
The {\it boundary measurement\/} between a given source  and  a given sink is then defined as the sum of path weights over all
(not necessary simple) paths between them. A boundary measurement is rational
 in the weights of edges and $\lambda$, see  ~\cite[Proposition~2.2]{GSV4}; in particular, if the network
 does not have oriented cycles then 
the boundary measurements are polynomials in edge weights, $\lambda$ and $\lambda^{-1}$.

 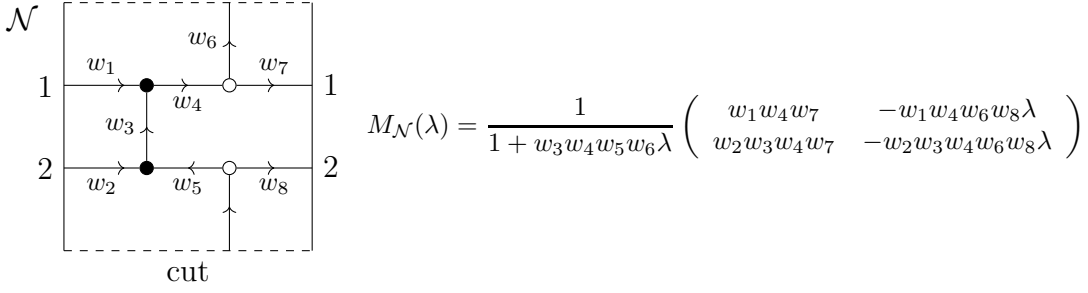
\begin{figure}[t]
 \centering
\begin{tikzpicture}[scale = 1.1]
\node [draw,circle,color=black, fill=black,inner sep=0pt,minimum size=5pt] (A) at (0,1) {};
\node [draw,circle,color=black, fill=white,inner sep=0pt,minimum size=5pt] (B) at (1,1) {};
\node [draw,circle,color=black, fill=black,inner sep=0pt,minimum size=5pt] (C) at (0,2) {};
\node [draw,circle,color=black, fill=white,inner sep=0pt,minimum size=5pt] (D) at (1,2) {};
\draw [-->-] (-1, 1) -- (A)   node[midway, below] {\footnotesize $ w_2$};
\draw [->-] (B) -- (A) node[midway, below] {\footnotesize $ w_5$};
\draw [->-] (A) -- (C) node[midway, left] {\footnotesize $ w_3$};
\draw [->-] (C) -- (D) node[midway, below] {\footnotesize $ w_4$};
\draw [-->-] (-1, 2) -- (C) node[midway, above] {\footnotesize $ w_1$};
\draw [->-] (B) -- (2,1) node[midway, below] {\footnotesize $ w_8$};
\draw [->-] (D) -- (2,2) node[midway, above] {\footnotesize $ w_7$};
\draw [->-] (D) -- +(0,1) node[midway, left] {\footnotesize $ w_6$};
\draw [-<-] (B) -- +(0,-1);
\draw [dashed] (-1,3) -- (2,3);
\draw (2,3) -- (2,0) node[pos = 0.33, right] {$1$} node[pos = 0.66, right] {$2$};;
\draw [ dashed] (-1,0) -- (2,0)  node[midway, below] {cut};;
\draw [ ] (-1,0) -- (-1,3)  node[pos = 0.33, left] {$2$} node[pos = 0.66, left] {$1$};;
\node () at (-1.5,2.8) {$\can$};
\node () at (7, 1.5) {\footnotesize $M_\can(\lambda) = \displaystyle\frac{1}{1 + w_3w_4w_5w_6\lambda}\left(\begin{array}{cc}w_1w_4w_7 & -w_1w_4w_6w_8  \lambda \\w_2w_3w_4w_7 & -w_2w_3w_4w_6w_8 \lambda\end{array}\right)$};
\end{tikzpicture}
\caption{A perfect network $\can$ on a cylinder and its boundary measurement matrix.}\label{fig:network}
\end{figure}

Boundary measurements are organized in a {\it boundary measurement matrix} $M_\can(\lambda)$, thus giving rise to the {\it boundary measurement map\/}
from $\EE_\can$ to the space of $n\times n$ rational matrix
functions, see e.g. Figure \ref{fig:network}. The \emph{gauge group} acts on $\EE_\can$ as follows: for any internal vertex $v$ of $\can$ and any $L\in \K^*$, the weights of all edges leaving $v$ are multiplied by $L$, and the weights of all edges entering $v$ are
 multiplied by $L^{-1}$. Clearly, the weights of paths between boundary vertices, and hence boundary measurements,
 are preserved under this action.
Therefore, the boundary measurement map factors through the space $\FF_\can$ defined as the
quotient of $\EE_\can$ by the action of the gauge group.

The quotient space $\FF_\can$ can be parametrized as follows.
The graph $\can$ divides $\mathcal C$ into a finite number of connected components called
\emph{faces}. The boundary of each face consists of edges of $\can$ and, possibly, of several arcs of
$\partial \mathcal C$.
A face is called {\it bounded\/} if its boundary contains only edges of $\can$ and {\it unbounded\/} otherwise.
Given a face $f$, we define its {\it face weight\/}
$y_f=\prod_{e\in\partial f}w_e^{\gamma_e}$,
where $\gamma_e=1$ if the direction of $e$ is compatible with the counterclockwise orientation of the
boundary $\partial f$ and $\gamma_e=-1$ otherwise.
Face weights are invariant under the gauge group action.
Then $\FF_\can$ is parametrized by the collection of all face weights (subject to condition $\prod_f y_f=1$)
 and a weight of an arbitrary path in $\can$ (not necessary directed) joining two boundary circles  (such a path is called a {\it trail\/}).
 
 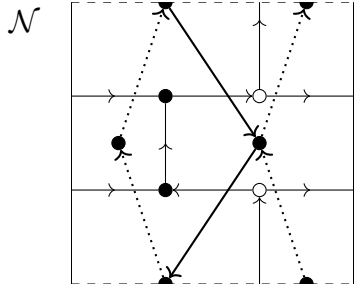
\begin{figure}[t]
 \centering
\begin{tikzpicture}[scale = 1.25]
\node [, opacity = 0.5] () at (-1.5,2.8) {$\can$};
\clip (-1,3) -- (-1,0) -- (2,0) -- (2,3) -- cycle;
\node [draw,circle,color=black, fill=black,inner sep=0pt,minimum size=5pt, opacity = 0.5] (A) at (0,1) {};
\node [draw,circle,color=black, fill=white,inner sep=0pt,minimum size=5pt, opacity = 0.5] (B) at (1,1) {};
\node [draw,circle,color=black, fill=black,inner sep=0pt,minimum size=5pt, opacity = 0.5] (C) at (0,2) {};
\node [draw,circle,color=black, fill=white,inner sep=0pt,minimum size=5pt, opacity = 0.5] (D) at (1,2) {};
\node [draw,circle,color=black, fill=black,inner sep=0pt,minimum size=5pt] (E) at (-0.5,1.5) {};
\node [draw,circle,color=black, fill=black,inner sep=0pt,minimum size=5pt] (F) at (0,3) {};
\node [draw,circle,color=black, fill=black,inner sep=0pt,minimum size=5pt] (G) at (0,0) {};
\node [draw,circle,color=black, fill=black,inner sep=0pt,minimum size=5pt] (H) at (1.5,0) {};
\node [draw,circle,color=black, fill=black,inner sep=0pt,minimum size=5pt] (I) at (1.5,3) {};
\node [draw,circle,color=black, fill=black,inner sep=0pt,minimum size=5pt] (J) at (1,1.5) {};
\draw [->, dotted, color = black, thick] (G) -- (E);
\draw [->, dotted, color = black, thick] (E) -- (F);
\draw [->, dotted, color = black, thick] (H) -- (J);
\draw [->, dotted, color = black, thick] (J) -- (I);
\draw [->, color = black, thick] (F) -- (J);
\draw [->, color = black, thick] (J) -- (G);
\draw [->-, opacity = 0.5] (-1, 1) -- (A);
\draw [->, opacity = 0.5] (B) -- (A);
\draw [->-, opacity = 0.5] (A) -- (C);
\draw [->, opacity = 0.5] (C) -- (D);
\draw [->-, opacity = 0.5] (-1, 2) -- (C);
\draw [->-, opacity = 0.5] (B) -- (2,1);
\draw [->-, opacity = 0.5] (D) -- (2,2); 
\draw [-->-, opacity = 0.5] (D) -- +(0,1);
\draw [<-, opacity = 0.5] (B) -- +(0,-1);
\draw [dashed] (-1,3) -- (2,3);
 \draw (2,3) -- (2,0);
\draw [ dashed] (-1,0) -- (2,0);
 \draw [ ] (-1,0) -- (-1,3);
\end{tikzpicture}
\caption{A perfect network $\can$ on a cylinder (grey) and its dual quiver $\Q_\can$ (black). Dotted arrows are half-edges.}\label{fig:quivernetwork}
\end{figure}

\paragraph{Poisson geometry of perfect cylindric networks.}
As  shown in \cite{GSV2, GSV4}, the space of edge weights $\EE_\can$ can be turned into a Poisson manifold
by considering Poisson brackets
that behave nicely with respect to a natural operation of concatenation of networks. Such Poisson brackets on $\EE_\can$ form a 6-parameter
family, which is pushed forward to a 2-parameter family of Poisson brackets on the quotient $\FF_\can$ of the edge weight space $\EE_\can$ by gauge transformations. We will pick a  specific member of the latter family. The corresponding Poisson structure, called {\em standard}, is described in terms of
the {\it quiver  $\Q_\can$ dual to the network\/} $\can$ (see Figure \ref{fig:quivernetwork}). Vertices of $\Q_\can$ are the faces of $\can$.
Edges of $\Q_\can$ correspond to the edges of $\can$ that connect either two internal vertices of different colors,
or an internal vertex with a boundary vertex. 
An edge $e^*$ in $Q_\can$ corresponding to $e$ in $\can$ is directed in such a way that the white endpoint of $e$ (if it exists) lies to the left of $e^*$ and
the black endpoint of $e$ (if it exists) lies to the right of $e$.
The weight $w^*(e^*)$ equals $1$  (i.e. $e^*$ is a conventional edge) if both endpoints of $e$ are internal vertices, and $1/2$  (i.e. $e^*$ is a half-edge) if one of the
endpoints of $e$ is a boundary vertex.  Then
the standard Poisson bracket on $\FF_\can$ descends to the space of face weights where it is given by
\begin{equation}
\label{facebracket}
\{y_f,y_{f'}\}=\left(\sum_{e^*: f\to f'} w^*(e^*)-
\sum_{e^*: f'\to f} w^*(e^*)\right)y_fy_{f'}.
\end{equation}
In other words, the bracket of face weights is the canonical Poisson bracket \eqref{eq:pb} associated with the quiver $\Q_\can$. As for the bracket of the trail weight  $z$ and a face weight $y_f$, its description in the general case is rather lengthy and will not be needed for our purposes.

\paragraph{Poisson property of the boundary measurement map and commuting Hamiltonians.}

To discuss Poisson properties of the boundary measurement map
defined above, we first  recall the definition of an R-matrix
(Sklyanin)
bracket, which plays a crucial role in the modern theory of integrable
systems, see e.g. \cite{FT, semenov2008research}. The bracket is defined on the space of $n\times n$ rational matrix
functions $M(\lambda)=(m_{ij}(\lambda) \in \K(\lambda))_{i,j=1}^n$ and is given by
the formula
\begin{equation}
\label{sklya}
\left \{M(\lambda){\stackrel{\textstyle{\small{\otimes}}}{,}} M(\mu)\right\}
= \left [ R(\lambda,\mu), M(\lambda)\otimes M(\mu) \right ],
\end{equation}
where the left-hand side is  understood as $$\left \{M(\lambda)
{\stackrel{\textstyle{\otimes}}{,}} M(\mu)\right\}_{ii'}^{jj'}=\{m_
{ij}(\lambda),m_{i'j'}(\mu)\}$$ and
the R-matrix $R(\lambda,\mu)$ is an operator in $\left(\R^n
\right )^{\otimes 2}$ depending on parameters $\lambda,\mu$ and solving
the classical Yang-Baxter equation. We are interested in the bracket
associated with the {\em trigonometric R-matrix}; its explicit form, which we will not need,
may be found e.g. in \cite[Section 4.5]{semenov2008research}.



\begin{theorem}
\label{Rtrig} {\bf{\cite{GSV4}}}
For any perfect network $\can$  on a cylinder, the above-defined Poisson structure on the space of edge weights induces the trigonometric R-matrix bracket on the space of boundary measurement matrices.

In other words, the boundary measurement map is a Poisson map from the space $\EE_\can$ of edge weights (and hence its quotient $\FF_\can$ by gauge transformations) to the space of $n\times n$ rational matrix
functions endowed with the trigonometric $R$-matrix bracket.
\end{theorem}

Furthermore, it is
well known that  {\em spectral invariants of an $n\times n$ rational matrix
function are in involution with respect to the Sklyanin bracket\/}, see \cite[Theorem 6.5]{semenov2008research}.  Therefore, one gets the following:
\begin{corollary}
For any perfect network $\can$  on a cylinder, spectral invariants of the boundary measurement matrix are Poisson-commuting functions on the space $\EE_\can$ of edge weights.
\end{corollary}
Furthermore, the so-defined commuting Hamiltonians descend to the quotient $\FF_\can$ by gauge transformations. In many examples the so-obtained family of functions in involution is an integrable system, and conversely, many integrable systems with an underlying cluster structure can be obtained in this way, see \cite{GSTV}.

\begin{remark}\label{pushforward}
One can also push the so-obtained integrable system forward to the cluster torus $Y_\Q$ associated with the dual quiver $\Q = \Q_\can$ of the network $\can$. More precisely, one gets an integrable system on the image $Y'_\Q := \{y_1y_2\dots = 1\}$ of the map $ \FF_\can \to Y_\Q$ which sends edge weights (modulo gauge transformations) to face weights. 
Indeed, although the boundary measurement map does not descend to $Y'_\Q$, boundary measurement matrices of edge weight collections belonging to the same fiber of the said map (i.e. having the same face weights) are related by the transformation $M(\lambda) \mapsto M(t \lambda)$, where $t \in \K^*$. Therefore, to get an integrable system on $Y'_\Q$ one can take central functions invariant under such transformations.
\end{remark}

\subsection{Discrete systems}
In this section we show that perfect cylindric networks also give rise to discrete integrable systems. Those systems are built out of \emph{Postnikov's moves} which are basically network versions of $Y$-type cluster mutations. In addition to those moves we will also need an operation that we call \emph{regluing}.

\begin{figure}[t]
\centering
\includegraphics[height=2.8in]{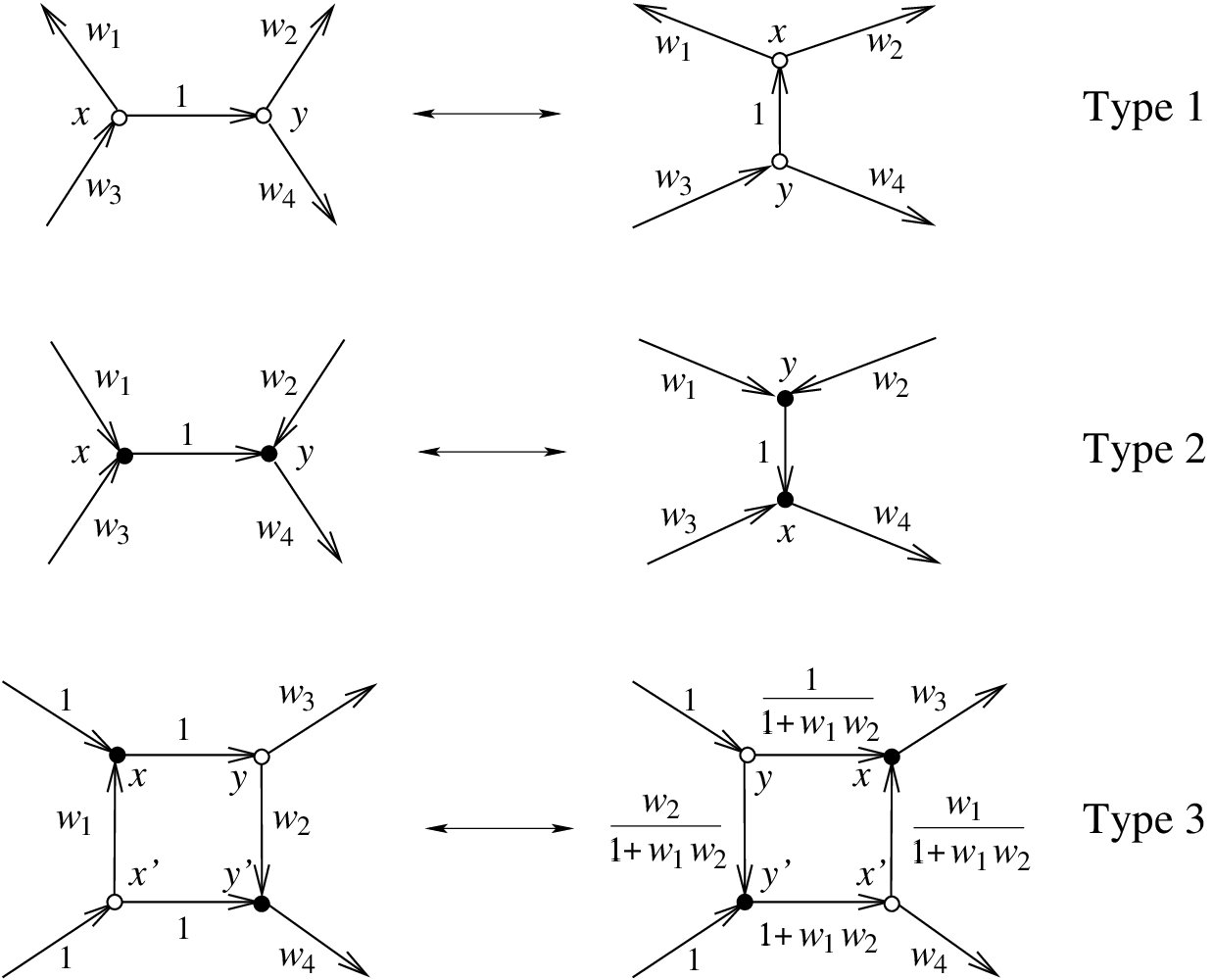}
\caption{Postnikov transformations }
\label{moves}
\end{figure}

\paragraph{Postnikov's moves.} Postnikov in \cite{Po}  introduced  elementary
transformations (or \emph{moves}) of weighted networks that do not change
the boundary measurement matrix. They are presented in Figure \ref{moves}. Each of those moves is preceded by a gauge transformation making the weights of indicated edges equal to $1$ and thus can be viewed as a map $\FF_\can \to \FF_{\can'}$, where $\can$ is the initial network and $\can'$ is the transformed one. Each such move is a Poisson map. Indeed, the moves of first two types do not change face or trail weights, while a Type 3 move (known as the \emph{square move}) acts on face weights as a $Y$-mutation of the dual quiver $\Q_\can$ at the vertex corresponding to the face where the move is being performed. Thus, if two networks $\can$ and $\tilde \can$ are related by a sequence of moves, then we have a Poisson map $\FF_\can \to \FF_{\tilde \can}$ identifying the corresponding boundary measurement matrices and hence integrable systems provided by their spectral invariants. 

 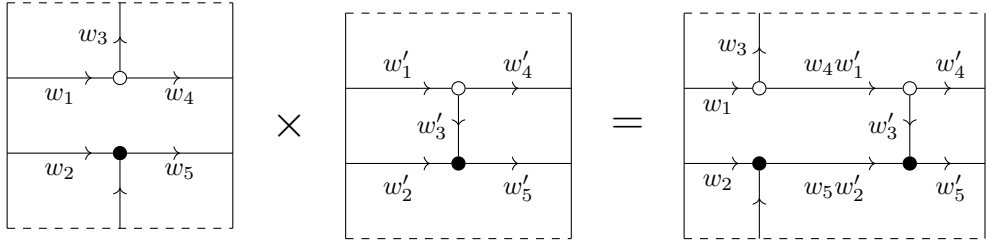
\begin{figure}[t]
 \centering
\begin{tikzpicture}[scale = 1]
 \node  () at (0,0) { \begin{tikzpicture}
\node [draw,circle,color=black, fill=black,inner sep=0pt,minimum size=5pt] (B) at (0.5,1) {};
\node [draw,circle,color=black, fill=white,inner sep=0pt,minimum size=5pt] (D) at (0.5,2) {};
\draw [-->-] (-1, 1) -- (B)   node[midway, below] {\footnotesize $ w_2$};
\draw [-->-] (-1, 2) -- (D) node[midway, below] {\footnotesize $ w_1$};
\draw [->-] (B) -- (2,1) node[midway, below] {\footnotesize $ w_5$};
\draw [->-] (D) -- (2,2) node[midway, below] {\footnotesize $ w_4$};
\draw [->-] (D) -- +(0,1) node[midway, left] {\footnotesize $ w_3$};
\draw [-<-] (B) -- +(0,-1);
\draw [dashed] (-1,3) -- (2,3);
\draw (2,3) -- (2,0); 
\draw [ dashed] (-1,0) -- (2,0)  node[midway, below] {};;
\draw [ ] (-1,0) -- (-1,3)  ; 
\end{tikzpicture}}; 
\node  () at (2.25,0) {\Large $\times$}; 
\node  () at (6.75,0) {\Large $=$}; 
 \node  () at (4.5,0) { \begin{tikzpicture}
\node [draw,circle,color=black, fill=black,inner sep=0pt,minimum size=5pt] (B) at (0.5,1) {};
\node [draw,circle,color=black, fill=white,inner sep=0pt,minimum size=5pt] (D) at (0.5,2) {};
\draw [-->-] (-1, 1) -- (B)   node[midway, below] {\footnotesize $ w_2'$};
\draw [-->-] (-1, 2) -- (D) node[midway, above] {\footnotesize $ w_1'$};
\draw [->-] (B) -- (2,1) node[midway, below] {\footnotesize $ w_5'$};
\draw [->-] (D) -- (2,2) node[midway, above] {\footnotesize $ w_4'$};
\draw [->-] (D) -- (B) node[midway, left] {\footnotesize $ w_3'$};
\draw [dashed] (-1,3) -- (2,3);
\draw (2,3) -- (2,0); 
\draw [ dashed] (-1,0) -- (2,0); 
\draw [ ] (-1,0) -- (-1,3); 
\end{tikzpicture}};
 \node  () at (9.5,0) { \begin{tikzpicture}
\node [draw,circle,color=black, fill=black,inner sep=0pt,minimum size=5pt] (B) at (0,1) {};
\node [draw,circle,color=black, fill=white,inner sep=0pt,minimum size=5pt] (D) at (0,2) {};
\draw [-->-] (-1, 1) -- (B)   node[midway, below] {\footnotesize $ w_2$};
\draw [-->-] (-1, 2) -- (D) node[midway, below] {\footnotesize $ w_1$};
\draw [->-] (D) -- +(0,1) node[midway, left] {\footnotesize $ w_3$};
\draw [-<-] (B) -- +(0,-1);
 
\node [draw,circle,color=black, fill=black,inner sep=0pt,minimum size=5pt] (B') at (2,1) {};
\node [draw,circle,color=black, fill=white,inner sep=0pt,minimum size=5pt] (D') at (2,2) {};
 \draw [-->-] (B) -- (B')   node[midway, below] {\footnotesize $ w_5\footnotesize w_2'$};
 \draw [-->-] (D) -- (D') node[midway, above] {\footnotesize $ w_4w_1'$};
\draw [->-] (B') -- (3,1) node[midway, below] {\footnotesize $ w_5'$};
\draw [->-] (D') -- (3,2) node[midway, above] {\footnotesize $ w_4'$};
\draw [->-] (D') -- (B') node[midway, left] {\footnotesize $ w_3'$};
\draw [dashed] (-1,3) -- (3,3);
\draw (3,3) -- (3,0); 
\draw [ dashed] (-1,0) -- (3,0) ; 
\draw [ ] (-1,0) -- (-1,3); 
\end{tikzpicture}};
\end{tikzpicture}
\caption{Concatenation of cylindric networks.}\label{fig:conc}
\end{figure}
\paragraph{Concatenation.} Suppose we have two cylindric networks $\can$, $\can'$, both with $n$ sources and sinks. Then we can identify the sinks of $\can$ with the corresponding sources of $\can'$, also gluing the corresponding boundaries of cylinders. The so-obtained cylindric network $\can \times \can'$ is called the \emph{concatenation} of $\can$ and $\can'$. The $2$-valent vertices of the concatenated network obtained by gluing sources and sinks are deleted, and the weight of each newly formed edge is defined to be the product of weights of  two corresponding edges of $\can$, $\can'$, see Figure \ref{fig:conc}. The boundary measurement matrix of the concatenation of two networks is the product of their boundary measurement matrices. Concatenation gives rise to Poisson maps $\EE_\can \times \EE_{\can'} \to \EE_{\can \times \can'}$ and $\FF_\can \times \FF_{\can'} \to \FF_{\can \times \can'}$.

\paragraph{Regluing.} Suppose we can represent a network $\can$ as concatenation of two networks $\can_1$, $\can_2$: $\can = \can_1 \times \can_2$. Note that edge weights of $\can_1$ and $\can_2$ are defined up to the following transformation: for each sink of $\can_1$, multiply the weight of the adjacent edge by a scalar; then divide the weight of the edge of $\can_2$ adjacent to the corresponding source by the same scalar. Now consider the network $\can' := \can_2 \times \can_1$ (i.e. concatenation of $\can_1$, $\can_2$ in the \emph{opposite order}). We call it the \emph{regluing} of the network $\can$. The weights of the reglued network are defined up to multiplying weights of edges adjacent to sources by scalars, and then dividing the weights of edges adjacent to sinks by the same scalars. Denote the group of such transformations by $T$. Then regluing can be viewed as a map $\FF_\can / T \to \FF_{\can'} / T$. From Poisson properties of concatenation it follows that this map is Poisson. Furthermore, boundary measurement matrices of $\can$ and $\can'$ are conjugate to each other: $M_{\can'} = M_{\can_2}M_{\can}M_{\can_2}^{-1}.$
So, just like moves, regluing preserves both the Poisson structure and the spectral invariants of the boundary measurement matrix.

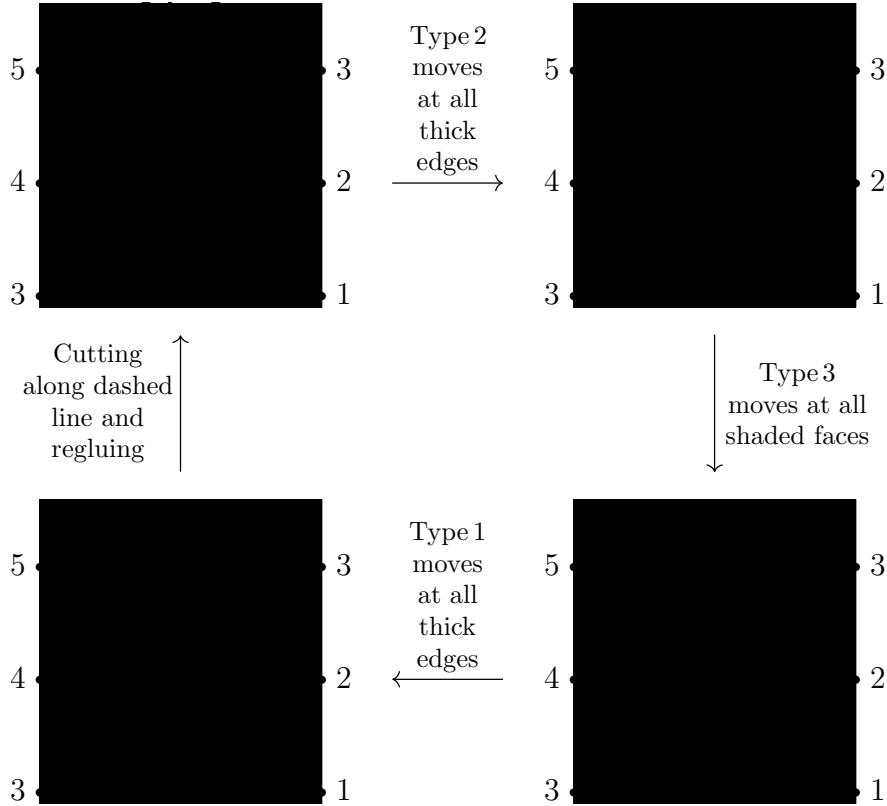
\begin{figure}[t]
\centering

\begin{tikzcd}[ row sep = large, column sep = large]
\begin{tikzpicture}[scale = 0.75]
\draw (0,-0.2) --  (0,5.2);
\clip (-1,-0.2) -- (6,-0.2) -- (6,5.2) -- (-1,5.2) -- cycle;
\fill [opacity = 0.05] (0,-0.2) -- (5,-0.2) -- (5,5.2) -- (0,5.2) -- cycle;
\draw [->-] (1,2) -- (2,3);
\draw [line width = 5, opacity = 0, shorten <= 20](1,2) -- (2,3);
\draw [line width = 5, opacity = 0, shorten <= 20](1,4) -- (2,3);
\draw [->-] (1,4) -- (2,3);
\draw [line width = 5, opacity = 0, shorten <= 20](1,4) -- (2,5);
\draw [->-] (1,4) -- (2,5);
\draw [line width = 5, opacity = 0, shorten <= 20](1,0) -- (2,1);
\draw [->-] (1,0) -- (2,1);
\draw [line width = 5, opacity = 0, shorten <= 20] (1,2) -- (2,1);;
\draw [->-] (1,2) -- (2,1);
\fill(0,0) circle (.4ex);
\fill(0,2) circle (.4ex);
\fill(0,4) circle (.4ex);
\draw [->-] (0,0) -- (1,0);
\draw [->-] (0,2) -- (1,2);
\draw [->-] (0,4) -- (1,4);
\fill(1,0) circle (.4ex);
\fill(1,2) circle (.4ex);
\fill(1,4) circle (.4ex);
\fill(2,1) circle (.4ex);
\fill(2,3) circle (.4ex);
\fill(2,5) circle (.4ex);
\draw [line width = 5, opacity = 0] (2,1) -- (3,1);
\draw [line width = 5, opacity = 0](2,3) -- (3,3);
\draw [line width = 5, opacity = 0] (2,5) -- (3,5);
\draw [->-, line width = 1.5] (2,1) -- (3,1);
\draw [->-, line width = 1.5] (2,3) -- (3,3);
\draw [->-, line width = 1.5] (2,5) -- (3,5);
\node at (0,0) [left] () {$3$};
\node at (0,2) [left] () {$4$};
\node at (0,4) [left] () {$5$};
\draw (1,0) -- (1.2,-0.2);
\draw (2,5) -- (1.8,5.2);
\draw [line width = 5, opacity = 0] (2,5) -- (1.8,5.2);
\draw (5,-0.2) -- (5,5.2);
\draw [-<-] (3,1) -- (4,2);
\draw [->-] (3,3) -- (4,2);
\draw [-<-] (3,3) -- (4,4);
\draw [->-] (3,1) -- (4,0);
\draw [->-] (3,5) -- (4,4);
\draw [line width = 5, opacity = 0, shorten >= 20](3,1) -- (4,2);
\draw [line width = 5, opacity = 0, shorten >= 20] (3,3) -- (4,2);
\draw [line width = 5, opacity = 0, shorten >= 20](3,3) -- (4,4);
\draw [line width = 5, opacity = 0, shorten >= 20](3,1) -- (4,0);
\draw [line width = 5, opacity = 0, shorten >= 20](3,5) -- (4,4);
\fill(3,1) circle (.4ex);
\fill(3,3) circle (.4ex);
\fill(4,0) circle (.4ex);
\fill(4,2) circle (.4ex);
\fill(4,4) circle (.4ex);
\fill(3,5) circle (.4ex);
\fill(5,0) circle (.4ex);
\fill(5,2) circle (.4ex);
\fill(5,4) circle (.4ex);
\draw [->-] (4,0) -- (5,0);
\draw [->-] (4,2) -- (5,2);
\draw [->-] (4,4) -- (5,4);
\node at (5,0) [right] () {$1$};
\node at (5,2) [right] () {$2$};
\node at (5,4) [right] () {$3$};
\draw (3,5) -- (3.2,5.2);
\draw (4,0) -- (3.8,-0.2);
\draw [line width = 5, opacity = 0](3,5) -- (3.2,5.2);
\end{tikzpicture} 
\arrow[yshift = 1.4cm]{r}{\parbox{1cm}{\centering \footnotesize {Type\,2 moves at all thick edges}}} &  \begin{tikzpicture}[scale = 0.75]
\clip (-1,-0.2) -- (6,-0.2) -- (6,5.2) -- (-1,5.2) -- cycle;
\draw (0,-0.2) --  (0,5.2);
\fill [opacity = 0.05] (0,-0.2) -- (5,-0.2) -- (5,5.2) -- (0,5.2) -- cycle;
\draw [->-] (1,2) -- (2.5,2.75);
\draw [->-] (1,4) -- (2.5,3.25);
\draw [->-] (1,4) -- (2.5,4.75);

\fill [opacity = 0.1] (2.5,4.75) -- (4,4) -- (2.5,3.25) -- (1,4) -- cycle;
\fill [opacity = 0.1] (2.5,2.75) -- (4,2) -- (2.5,1.25) -- (1,2) -- cycle;
\fill [opacity = 0.1] (2.5,0.75) -- (4,0) -- (2.5,-0.75) -- (1,0) -- cycle;
\draw [->-] (1,0) -- (2.5,0.75);
\draw [->-] (1,2) -- (2.5,1.25);
\fill(0,0) circle (.4ex);
\fill(0,2) circle (.4ex);
\fill(0,4) circle (.4ex);
\draw [->-] (0,0) -- (1,0);
\draw [->-] (0,2) -- (1,2);
\draw [->-] (0,4) -- (1,4);
\fill(1,0) circle (.4ex);
\fill(1,2) circle (.4ex);
\fill(1,4) circle (.4ex);
\fill(2.5,0.75) circle (.4ex);
\fill(2.5,3.25) circle (.4ex);
\draw [->-] (2.5,1.25) -- (2.5,0.75);
\draw [->-] (2.5,3.25) -- (2.5,2.75);
\draw [->-] (2.5,5.2) -- (2.5,4.75);
\node at (0,0) [left] () {$3$};
\node at (0,2) [left] () {$4$};
\node at (0,4) [left] () {$5$};
\draw (1,0) -- (1.4,-0.2);
\draw (5,-0.2) -- (5,5.2);
\draw [-<-] (2.5,1.25) -- (4,2);
\draw [->-] (2.5,2.75) -- (4,2);
\draw [-<-] (2.5,3.25) -- (4,4);
\draw [->-] (2.5,0.75) -- (4,0);
\draw [->-] (2.5,4.75) -- (4,4);
\fill(2.5,1.25) circle (.4ex);
\fill(2.5,2.75) circle (.4ex);
\fill(4,0) circle (.4ex);
\fill(4,2) circle (.4ex);
\fill(4,4) circle (.4ex);
\fill(2.5,4.75) circle (.4ex);
\fill(5,0) circle (.4ex);
\fill(5,2) circle (.4ex);
\fill(5,4) circle (.4ex);
\draw [->-] (4,0) -- (5,0);
\draw [->-] (4,2) -- (5,2);
\draw [->-] (4,4) -- (5,4);
\node at (5,0) [right] () {$1$};
\node at (5,2) [right] () {$2$};
\node at (5,4) [right] () {$3$};
\draw (4,0) -- (3.6,-0.2);

\end{tikzpicture} \arrow[shorten >=0.2cm,shorten <=0.2cm]{dd}{\parbox{2cm}{\centering \footnotesize Type\,3 moves at all shaded faces}}\\
\\
\begin{tikzpicture}[scale = 0.75]
\draw (0,-0.2) --  (0,5.2);
\clip (-1,-0.2) -- (6,-0.2) -- (6,5.2) -- (-1,5.2) -- cycle;
\fill [opacity = 0.05] (0,-0.2) -- (5,-0.2) -- (5,5.2) -- (0,5.2) -- cycle;
\draw [-<-] (1,2) -- (2,3);
\draw [line width = 5, opacity = 0, shorten <= 20](1,2) -- (2,3);

\draw [line width = 5, opacity = 0, shorten <= 20](1,4) -- (2,3);
\draw [->-] (1,4) -- (2,3);
\draw [line width = 5, opacity = 0, shorten <= 20](1,4) -- (2,5);
\draw [-<-] (1,4) -- (2,5);
\draw [line width = 5, opacity = 0, shorten <= 20](1,0) -- (2,1);
\draw [-<-] (1,0) -- (2,1);
\draw [line width = 5, opacity = 0, shorten <= 20] (1,2) -- (2,1);;
\draw [->-] (1,2) -- (2,1);
\fill(0,0) circle (.4ex);
\fill(0,2) circle (.4ex);
\fill(0,4) circle (.4ex);
\draw [->-] (0,0) -- (1,0);
\draw [->-] (0,2) -- (1,2);
\draw [->-] (0,4) -- (1,4);
\fill(1,0) circle (.4ex);
\fill(1,2) circle (.4ex);
\fill(1,4) circle (.4ex);
\fill(2,1) circle (.4ex);
\fill(2,3) circle (.4ex);
\fill(2,5) circle (.4ex);
\draw [-->-] (2,1) -- (3,1);
\draw [-->-] (2,3) -- (3,3);
\draw [-->-] (2,5) -- (3,5);
\node at (0,0) [left] () {$3$};
\node at (0,2) [left] () {$4$};
\node at (0,4) [left] () {$5$};
\draw [dashed] (2.5,-0.2) -- (2.5,5.2);
\draw (1,0) -- (1.2,-0.2);
\draw (2,5) -- (1.8,5.2);
\draw [line width = 5, opacity = 0] (2,5) -- (1.8,5.2);
\draw (5,-0.2) -- (5,5.2);
\draw [->-] (3,1) -- (4,2);
\draw [->-] (3,3) -- (4,2);
\draw [->-] (3,3) -- (4,4);
\draw [->-] (3,1) -- (4,0);
\draw [->-] (3,5) -- (4,4);

\fill(3,1) circle (.4ex);
\fill(3,3) circle (.4ex);
\fill(4,0) circle (.4ex);
\fill(4,2) circle (.4ex);
\fill(4,4) circle (.4ex);
\fill(3,5) circle (.4ex);
\fill(5,0) circle (.4ex);
\fill(5,2) circle (.4ex);
\fill(5,4) circle (.4ex);
\draw [->-] (4,0) -- (5,0);
\draw [->-] (4,2) -- (5,2);
\draw [->-] (4,4) -- (5,4);
\node at (5,0) [right] () {$1$};
\node at (5,2) [right] () {$2$};
\node at (5,4) [right] () {$3$};
\node at (2.3,1) [below] () {$3$};
\node at (2.3,3) [below] () {$4$};
\node at (2.3,5) [below] () {$5$};

\draw (3,5) -- (3.2,5.2);
\draw (4,0) -- (3.8,-0.2);

\draw [line width = 5, opacity = 0](3,5) -- (3.2,5.2);


\end{tikzpicture}\arrow[shorten >=0.2cm,shorten <=0.2cm]{uu}{\parbox{2cm}{\centering \footnotesize Cutting along dashed line and regluing}}    &  \begin{tikzpicture}[scale = 0.75]
\clip (-1,-0.2) -- (6,-0.2) -- (6,5.2) -- (-1,5.2) -- cycle;
\draw (0,-0.2) --  (0,5.2);
\fill [opacity = 0.05] (0,-0.2) -- (5,-0.2) -- (5,5.2) -- (0,5.2) -- cycle;
\draw [-<-] (1,2) -- (2.5,2.75);
\draw [->-] (1,4) -- (2.5,3.25);
\draw [-<-] (1,4) -- (2.5,4.75);

\draw [-<-] (1,0) -- (2.5,0.75);
\draw [->-] (1,2) -- (2.5,1.25);
\fill(0,0) circle (.4ex);
\fill(0,2) circle (.4ex);
\fill(0,4) circle (.4ex);
\draw [->-] (0,0) -- (1,0);
\draw [->-] (0,2) -- (1,2);
\draw [->-] (0,4) -- (1,4);
\fill(1,0) circle (.4ex);
\fill(1,2) circle (.4ex);
\fill(1,4) circle (.4ex);
\fill(2.5,0.75) circle (.4ex);
\fill(2.5,3.25) circle (.4ex);
\draw [->, line width = 1.5] (2.5,1.25) -- (2.5,0.75);
\draw [->,  line width = 1.5] (2.5,3.25) -- (2.5,2.75);
\draw [->,  line width = 1.5] (2.5,5.2) -- (2.5,4.75);
\node at (0,0) [left] () {$3$};
\node at (0,2) [left] () {$4$};
\node at (0,4) [left] () {$5$};
\draw (1,0) -- (1.4,-0.2);
\draw (5,-0.2) -- (5,5.2);
\draw [->-] (2.5,1.25) -- (4,2);
\draw [->-] (2.5,2.75) -- (4,2);
\draw [->-] (2.5,3.25) -- (4,4);
\draw [->-] (2.5,0.75) -- (4,0);
\draw [->-] (2.5,4.75) -- (4,4);
\fill(2.5,1.25) circle (.4ex);
\fill(2.5,2.75) circle (.4ex);
\fill(4,0) circle (.4ex);
\fill(4,2) circle (.4ex);
\fill(4,4) circle (.4ex);
\fill(2.5,4.75) circle (.4ex);
\fill(5,0) circle (.4ex);
\fill(5,2) circle (.4ex);
\fill(5,4) circle (.4ex);
\draw [->-] (4,0) -- (5,0);
\draw [->-] (4,2) -- (5,2);
\draw [->-] (4,4) -- (5,4);
\node at (5,0) [right] () {$1$};
\node at (5,2) [right] () {$2$};
\node at (5,4) [right] () {$3$};
\draw (4,0) -- (3.6,-0.2);

\end{tikzpicture}
\arrow[yshift = 1.4cm, swap]{l}{\parbox{1.2cm}{\centering \footnotesize Type\,1 moves at all thick edges}} 
\end{tikzcd}.

\caption{Network representation of the pentagram map.}\label{FigPSIDO}
\end{figure}

\paragraph{Construction of discrete integrable systems.}

Now assume that after a sequence of moves and  regluing  a network $\can$ transforms into a network $\can'$ which is isomorphic to $\can$ as a graph on a cylinder (i.e. we have an isomorphism $\can \to \can'$ of directed graphs which extends to an orientation-preserving self-homeomorphism of the cylinder). Then we obtain a Poisson self-map $\FF_\can / T \to \FF_\can / T$ which preserves the spectral invariants of the boundary measurement matrix, i.e. is a discrete integrable system (assuming that the spectral invariants themselves define a continuous integrable system). For instance, Figure \ref{FigPSIDO} gives a network representation of one of the best known cluster integrable systems: the pentagram map (see Section \ref{sec:further}).

One can interpret such discrete integrable systems obtained from network moves and regluings as cluster maps. The corresponding quiver is dual to the network on the torus obtained from the initial cylindric network by gluing sources to the corresponding sinks.


\section{Cluster integrable systems from dimers}\label{sec:GK}
In this section we describe the Goncharov-Kenyon construction of cluster integrable systems based on the dimer model \cite{GK}.

\subsection{Continuous systems}\label{sec:cont}

\paragraph{The dimer model.} Given an undirected graph, its \textit{dimer cover} (or a \textit{perfect matching}) is a set of edges with the property that every vertex is adjacent to a unique edge of the cover. A \textit{weighted graph} is a graph with non-zero numbers (i.e. elements of $\K^*$ where $\K$ is the base field of real or complex numbers) assigned to edges. Given a dimer cover of a weighted graph $\Gamma$, the weight of that cover is defined as the product of weights of its edges. The study of perfect matchings of a given (weighted) graph is known as the \emph{dimer model}. In this paper we are only interested in the dimer model on \emph{bipartite graphs}, i.e. graphs whose vertices are colored black and white in such a way that each edge has one white vertex and one black vertex. 


\paragraph{Dimer models on surface graphs.}

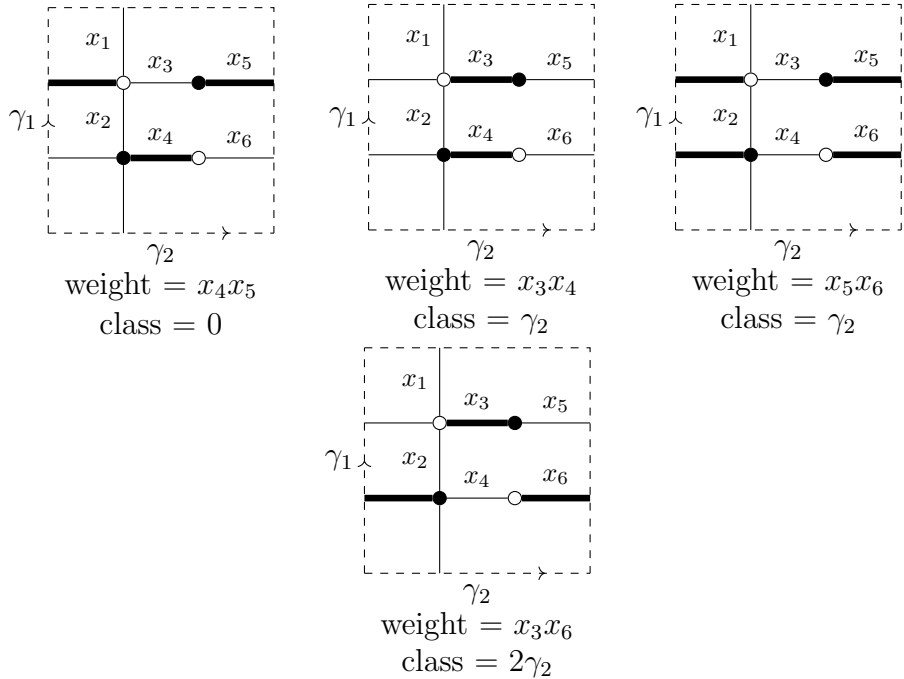
\begin{figure}[t]
 \centering
 \begin{tikzpicture}
\node [draw,circle,color=black, fill=black,inner sep=0pt,minimum size=5pt] (B) at (0,1) {};
\node [draw,circle,color=black, fill=white,inner sep=0pt,minimum size=5pt] (C) at (0,2) {};
\node [draw,circle,color=black, fill=white,inner sep=0pt,minimum size=5pt] (B1) at (1,1) {};
\node [draw,circle,color=black, fill=black,inner sep=0pt,minimum size=5pt] (C1) at (1,2) {};

\draw [line width = 2.5pt]  (B) -- (B1) node[midway, above] {\footnotesize  $x_4$};
\draw(C) -- (C1)  node[midway, above] {\footnotesize $x_3$};
\draw  (C) -- (B)  node[midway, left] {\footnotesize $x_2$};
\draw (C) -- +(0,1)  node[midway, left] {\footnotesize $x_1$};
\draw  (B) -- +(0,-1);
\draw (B) -- +(-1,0) ;
\draw  [line width = 2.5pt]  (C) -- +(-1,0) ;
\draw  (B1) -- +(1,0)   node[midway, above] {\footnotesize $x_6$};
\draw  [line width = 2.5pt]   (C1) -- +(1,0)  node[midway, above] {\footnotesize $x_5$};
\draw [dashed] (-1,3) -- (2,3) -- (2,0);
\draw [-->-, dashed] (-1,0) -- (2,0)  node[midway, below] {$\gamma_2$};;
\draw [->-, dashed] (-1,0) -- (-1,3)  node[midway, left] {$\gamma_1$};;
\node  () at (0.5,-0.7) {weight = $x_4x_5$};
\node  () at (0.5,-1.2) {class = $0$};
\end{tikzpicture} \, \,
\begin{tikzpicture}
\node  () at (0.5,-0.7) {weight = $x_3x_4$};
\node  () at (0.5,-1.2) {class = $\gamma_2$};
\node [draw,circle,color=black, fill=black,inner sep=0pt,minimum size=5pt] (B) at (0,1) {};
\node [draw,circle,color=black, fill=white,inner sep=0pt,minimum size=5pt] (C) at (0,2) {};
\node [draw,circle,color=black, fill=white,inner sep=0pt,minimum size=5pt] (B1) at (1,1) {};
\node [draw,circle,color=black, fill=black,inner sep=0pt,minimum size=5pt] (C1) at (1,2) {};

\draw [line width = 2.5pt]  (B) -- (B1) node[midway, above] {\footnotesize $x_4$};
\draw [line width = 2.5pt] (C) -- (C1)  node[midway, above] {\footnotesize $x_3$};
\draw  (C) -- (B)  node[midway, left] {\footnotesize $x_2$};
\draw (C) -- +(0,1)  node[midway, left] {\footnotesize $x_1$};
\draw  (B) -- +(0,-1);
\draw (B) -- +(-1,0) ;
\draw (C) -- +(-1,0)  ;
\draw  (B1) -- +(1,0)  node[midway, above] {\footnotesize $x_6$} ;
\draw  (C1) -- +(1,0) node[midway, above] {\footnotesize $x_5$};
\draw [dashed] (-1,3) -- (2,3) -- (2,0);
\draw [-->-, dashed] (-1,0) -- (2,0)  node[midway, below] {$\gamma_2$};;
\draw [->-, dashed] (-1,0) -- (-1,3)  node[midway, left] {$\gamma_1$};;
\end{tikzpicture} \,\,
\begin{tikzpicture}
\node  () at (0.5,-0.7) {weight = $x_5x_6$};
\node  () at (0.5,-1.2) {class = $\gamma_2$};
\node [draw,circle,color=black, fill=black,inner sep=0pt,minimum size=5pt] (B) at (0,1) {};
\node [draw,circle,color=black, fill=white,inner sep=0pt,minimum size=5pt] (C) at (0,2) {};
\node [draw,circle,color=black, fill=white,inner sep=0pt,minimum size=5pt] (B1) at (1,1) {};
\node [draw,circle,color=black, fill=black,inner sep=0pt,minimum size=5pt] (C1) at (1,2) {};

\draw (B) -- (B1) node[midway, above] {\footnotesize $x_4$};
\draw (C) -- (C1)  node[midway, above] {\footnotesize $x_3$};
\draw  (C) -- (B)  node[midway, left] {\footnotesize $x_2$};
\draw (C) -- +(0,1)  node[midway, left] {\footnotesize $x_1$};
\draw  (B) -- +(0,-1);
\draw  [line width = 2.5pt]   (B) -- +(-1,0)  ;
\draw[line width = 2.5pt]  (C) -- +(-1,0)  ;
\draw  [line width = 2.5pt]   (B1) -- +(1,0)  node[midway, above] {\footnotesize $x_6$};
\draw [line width = 2.5pt]   (C1) -- +(1,0) node[midway, above] {\footnotesize $x_5$};
\draw [dashed] (-1,3) -- (2,3) -- (2,0);
\draw [-->-, dashed] (-1,0) -- (2,0)  node[midway, below] {$\gamma_2$};;
\draw [->-, dashed] (-1,0) -- (-1,3)  node[midway, left] {$\gamma_1$};;
\end{tikzpicture}\,\,
\begin{tikzpicture}
\node  () at (0.5,-0.7) {weight = $x_3x_6$};
\node  () at (0.5,-1.2) {class = $2\gamma_2$};
\node [draw,circle,color=black, fill=black,inner sep=0pt,minimum size=5pt] (B) at (0,1) {};
\node [draw,circle,color=black, fill=white,inner sep=0pt,minimum size=5pt] (C) at (0,2) {};
\node [draw,circle,color=black, fill=white,inner sep=0pt,minimum size=5pt] (B1) at (1,1) {};
\node [draw,circle,color=black, fill=black,inner sep=0pt,minimum size=5pt] (C1) at (1,2) {};

\draw (B) -- (B1) node[midway, above] {\footnotesize $x_4$};
\draw [line width = 2.5pt] (C) -- (C1)  node[midway, above] {\footnotesize $x_3$};
\draw  (C) -- (B)  node[midway, left] {\footnotesize $x_2$};
\draw (C) -- +(0,1)  node[midway, left] {\footnotesize $x_1$};
\draw  (B) -- +(0,-1);
\draw  [line width = 2.5pt]   (B) -- +(-1,0)  ;
\draw (C) -- +(-1,0)  ;
\draw  [line width = 2.5pt]   (B1) -- +(1,0) node[midway, above] {\footnotesize $x_6$} ;
\draw  (C1) -- +(1,0) node[midway, above] {\footnotesize $x_5$};
\draw [dashed] (-1,3) -- (2,3) -- (2,0);
\draw [-->-, dashed] (-1,0) -- (2,0)  node[midway, below] {$\gamma_2$};;
\draw [->-, dashed] (-1,0) -- (-1,3)  node[midway, left] {$\gamma_1$};;
\end{tikzpicture}
 
\caption{Dimer covers of a bipartite graph on a torus, along with their weights and homology classes.}\label{fig:dt}
\end{figure}

A \textit{surface graph} $\Gamma$ is a graph embedded in a $2$-dimensional surface $\Sigma$ in such a way that its \textit{faces}, i.e.  connected components of the complement $\Sigma \setminus \Gamma$, are contractible.  
For surface bipartite graphs one can assign additional information to dimer covers, namely \emph{homology classes}.
Let $\Gamma$ be a surface bipartite graph on a surface $\Sigma$. Then, since the edges of $\Gamma$ can be canonically oriented from white to black, any dimer cover may be viewed as an integral $1$-chain. Furthermore, all such chains have the same boundary, namely the sum of black vertices minus the sum of white vertices. In other words, the difference of two dimer covers is a cycle. Fixing a reference dimer cover $D_0$, one defines the homology class of any other dimer cover $D$ as the class of $D-D_0$ in $H_1(\Sigma; \Z)$. Figure \ref{fig:dt} shows a weighted bipartite graph on a torus, its dimer covers, their weights, and their homology classes. Here the reference dimer cover is the leftmost one.

\paragraph{Goncharov-Kenyon Hamiltonians.}
The Goncharov-Kenyon integrable system is defined based on a surface bipartite graph on a torus, $\Sigma = T^2$.
The \textit{Goncharov-Kenyon Hamiltonian} $H_\xi$ corresponding to a class $\xi \in H_1(T^2; \Z)$ is defined as the sum of weights of all dimer covers in the class $\xi$. Those Hamiltonians are considered as functions on the space of edge weights up to \textit{gauge transformations}. A gauge transformation is multiplication of weights of all edges adjacent to a given vertex by a given number $\lambda \in \K^*$. Viewing the space of edge weights as the space of $1$-cochains, one identifies its quotient by gauge transformations with the cohomology group $H^1(\Gamma; \K^*)$. Since a gauge transformation multiplies all Goncharov-Kenyon Hamiltonians $H_\xi$ by the same number, the Hamiltonians are well-defined as functions on $H^1(\Gamma; \K^*)$, up to a common factor.  

\begin{example}
For the graph in Figure \ref{fig:dt}, the Goncharov-Kenyon Hamiltonians are
$
H_0 = x_4x_5$, $H_{\gamma_2} = x_3x_4 + x_5x_6$, $H_{2\gamma_2} = x_3x_6.
$
The quotient of any two of these Hamiltonians is a well-defined function on $H^1(\Gamma; \K^*)$. For instance,
$
{H_{\gamma_2}} {H_0}^{-1} = {x_3}{x_5}^{-1} + {x_6}{x_4}^{-1} = \langle {e_3 - e_5}\rangle + \langle{e_6 - e_4}\rangle,
$
where $e_i$ is the edge with label $x_i$ (oriented from white to black), and $\langle a \rangle$ is a function on $H^1(\Gamma; \K^*)$ given by pairing with a cycle $a \in H_1(\Gamma; \Z)$.
\end{example}


\paragraph{Surface graphs and Poisson structures.} \begin{figure}[b]
 \centering
\begin{tikzpicture}[]
\node [draw,circle,color=black, fill=black,inner sep=0pt,minimum size=5pt] (B) at (0,0.8) {};
\node [draw,circle,color=black, fill=white,inner sep=0pt,minimum size=5pt] (C) at (0,2.2) {};
\node [draw,circle,color=black, fill=white,inner sep=0pt,minimum size=5pt] (B1) at (1.4,0.8) {};
\node [draw,circle,color=black, fill=black,inner sep=0pt,minimum size=5pt] (C1) at (1.4,2.2) {};
\draw  (B) -- (B1);
\draw (C) -- (C1);
\draw (C) -- (B);
\draw (C1) -- (B1);
\draw  (C) -- +(0,1);
\draw  (B) -- +(0,-1);
\draw  (B) -- +(-1,0)  ;
\draw  (C) -- +(-1,0) ;
\draw  (B1) -- +(1,0);
\draw  (C1) -- +(1,0) ;
\draw  (B1) -- +(0,-1);
\draw  (C1) -- +(0,1) ;
\fill [opacity = 0.1] (-0.2, -0.2) -- (0.2,-0.2) -- (0.2,3.2) -- (-0.2,3.2) -- cycle;
\draw [thick, densely dotted] (-0.2, 0.1) -- (0.2,0.1);
\draw [thick, densely dotted] (-0.2, 2.9) -- (0.2,2.9);
\draw [thick, densely dotted] (-0.2, 1.5) -- (0.2,1.5);
\fill [opacity = 0.1] (1.2, -0.2) -- (1.6,-0.2) -- (1.6,3.2) -- (1.2,3.2) -- cycle;
\draw [thick, densely dotted] (1.2, 0.1) -- (1.6,0.1);
\draw [thick, densely dotted] (1.2, 2.9) -- (1.6,2.9);
\draw [thick, densely dotted] (1.2, 1.5) -- (1.6,1.5);
\fill [opacity = 0.1] (-1, 0.6) -- (-1,1) -- (2.4,1) -- (2.4, 0.6) -- cycle;
\draw [thick, densely dotted] (-0.7, 0.6) -- (-0.7,1);
\draw [thick, densely dotted] (2.1, 0.6) -- (2.1,1);
\draw [thick, densely dotted] (0.7, 0.6) -- (0.7,1);
\fill [opacity = 0.1] (-1, 2) -- (-1,2.4) -- (2.4,2.4) -- (2.4, 2) -- cycle;
\draw [thick, densely dotted] (-0.7, 2) -- (-0.7,2.4);
\draw [thick, densely dotted] (2.1, 2) -- (2.1,2.4);
\draw [thick, densely dotted] (0.7, 2) -- (0.7,2.4);
\end{tikzpicture}
\caption{A surface bipartite graph with its tubular neighborhood $S$. The conjugate surface $\tilde S$ is defined by cutting $S$ along dotted lines and then reattaching the sides of each cut with opposite orientations. }\label{fig:conj}
\end{figure}
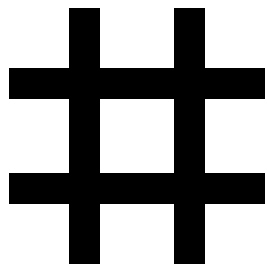 Given a surface bipartite graph $\Gamma$ on an orientable surface $\Sigma$, one defines a Poisson structure on $H^1(\Gamma; \K^*)$ as follows. Let $S$ be a tubular neighborhood of $\Gamma$ in $\Sigma$ which retracts to $\Gamma$. The \emph{conjugate surface} $\tilde S$ is defined by cutting $S$ along perpendicular bisectors of edges of $\Gamma$ and then reattaching the sides of each cut with opposite orientations (see Figure~\ref{fig:conj}). This new surface $\tilde S$ is still orientable, thanks to the graph $\Gamma$ being bipartite. So, we have an intersection pairing in $H_1(\tilde S; \Z)$, defined up to sign. Furthermore, the graph $\Gamma$ is a deformation retract of $\tilde S$, giving an intersection pairing in $H_1(\Gamma; \Z)$ as well. This gives a Poisson structure on $H^1(\Gamma; \K^*)$ defined as follows. Let $\langle a \rangle$ be a function on $H^1(\Gamma; \K^*)$ defined by pairing with a cycle  $a \in H_1(\Gamma; \Z)$. One can use such functions to define coordinates on $H^1(\Gamma; \K^*)$, so it is sufficient to describe Poisson brackets for functions of that form. Those brackets are given by
$
\{\langle a \rangle, \langle b \rangle\} = (a \cap b)  \langle a \rangle \langle b \rangle,
$
where $a \cap b$ is the intersection number.

The so-obtained Poisson structure on $H^1(\Gamma; \K^*)$ can be interpreted as an ``extension'' of a cluster Poisson structure. Since the graph $\Gamma$ is bipartite, it has a canonical orientation (from white to black). Take its oriented dual (in $\Sigma$) and delete all cycles of length $2$ (do it just like for quiver mutations: one at a time, until there are no such cycles left). This gives a quiver $\mathcal Q$ embedded in $\Sigma$, called the \emph{dual quiver} of $\Gamma$. Furthermore, we have a map $\delta \colon H^1(\Gamma; \K^*) \to Y_\Q$ defined as follows: the image of a cocycle $\alpha$ is a function that assigns to each vertex of $\Q$ the value of $\alpha$ on the boundary of the corresponding face of $\Gamma$. One can easily see from the definition of the conjugate surface that
the map $\delta \colon H^1(\Gamma; \K^*) \to Y_\Q$ is Poisson (up to sign). One can think of this map $\delta$ as the quotient map for the natural action of $H^1(\Sigma, \K^*)$ on $H^1(\Gamma; \K^*)$. The image of $\delta$ is a Poisson submanifold of $Y_\Q$ consisting of functions whose values at all vertices of $\Q$ multiply to $1$.


\par

\paragraph{Continuous dimer integrable systems.}
The main result of \cite{GK} concerns \emph{minimal} bipartite graphs on the torus. The definition of a minimal graph is quite technical and will not be discussed here. For minimal graphs the set of all $\xi \in H_1(T^2; \Z)$ such that $H_\xi \neq 0$ coincides with the set of lattice points in a certain convex polygon, called the \emph{Newton polygon} of $\Gamma$.

\begin{theorem} {\bf{\cite{GK}}}
For a minimal\footnote{As follows from \cite{izosimov2022dimers}, Hamiltonians $H_\xi$ Poisson commute for any toric bipartite graph, minimal or not. In the non-minimal case the number of those Hamiltonians might be not maximal (so that we do not obtain an integrable system).} bipartite graph $\Gamma$ on a torus, normalized\footnote{A normalization is needed since Goncharov-Kenyon Hamiltonians are only defined up to a common factor. For instance, one can normalize by choosing one Goncharov-Kenyon Hamiltonian $H$ and considering quotients $H_\xi / H$.} Goncharov-Kenyon Hamiltonians $H_\xi$ form a completely integrable system on the Poisson manifold $H^1(\Gamma; \K^*)$. Furthermore, under a suitable normalization, Hamiltonians corresponding to boundary points of the Newton polygon are exactly the Casimirs, while Hamiltonians corresponding to interior points are independent on the symplectic leaves.
\end{theorem}

   
 \begin{remark}[cf. Remark \ref{pushforward}]
It is also common to interpret Goncharov-Kenyon Hamiltonians as an integrable system on the cluster torus $Y_\Q$ associated with the dual quiver $\Q$ of the graph $\Gamma$, cf. \cite{FM}. More precisely, one gets an integrable system on the image $Y'_\Q := \{y_1y_2\dots = 1\}$ of the map $\delta \colon H^1(\Gamma; \K^*) \to Y_\Q$. Indeed, recall that $\delta$ is the quotient map for the action of $H^1(T^2; \K^*)$ on $H^1(\Gamma; \K^*)$. Under the action of a cohomology class $\alpha \in H^1(T^2; \K^*)$, the Goncharov-Kenyon Hamiltonian $H_\xi$, where $\xi \in H_1(T^2; \Z)$, transforms by the rule $H_{\xi} \mapsto \alpha(\xi) H_\xi$. Any monomial of Goncharov-Kenyon Hamiltonians invariant under such transformations (as well as under simultaneous rescaling of all Hamiltonians) descends to the image $Y_\Q'$ of~$\delta$. Taking all such monomials one gets an integrable system on $Y_\Q'$. 
    
 \end{remark}

\subsection{Discrete systems}\label{sec:ds}
Let $\Gamma$ be a toric bipartite graph, as in Section \ref{sec:cont} above. 
\emph{Discrete Goncharov-Kenyon systems} are Poisson automorphisms of $H^1(\Gamma; \K^*)$ which preserve Goncharov-Kenyon Hamiltonians. Those automorphisms are built out of \emph{local moves} which are dimer model versions of Postnikov's moves and hence $Y$-type cluster mutations.

 \begin{figure}[t]
 \centering
\begin{tikzpicture}[, scale = 0.75]
\node [draw,circle,color=black, fill=black,inner sep=0pt,minimum size=5pt] (A) at (0,0) {};
\node [draw,circle,color=black, fill=white,inner sep=0pt,minimum size=5pt] (B) at (1,0) {};
\node [draw,circle,color=black, fill=black,inner sep=0pt,minimum size=5pt] (C) at (2,0) {};
\node [draw,circle,color=black, fill=black,inner sep=0pt,minimum size=5pt] (D) at (6,0) {};
\draw (A) -- (B) -- (C);
\draw (A) -- +(-0.5,+0.5);
\draw (A) -- +(-0.5,-0.5);
\draw (C) -- +(+0.5,+0.5);
\draw (C) -- +(+0.5,-0.5);
\node () at (4,0) {$\longleftrightarrow$};
\draw (D) -- +(-0.5,+0.5);
\draw (D) -- +(-0.5,-0.5);
\draw (D) -- +(+0.5,+0.5);
\draw (D) -- +(+0.5,-0.5);

\end{tikzpicture}
\caption{Shrinking a $2$-valent vertex.}\label{shrink}
\end{figure}
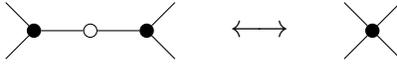
\paragraph{Shrinking of a $2$-valent vertex.} The first move type is \emph{shrinking of a $2$-valent vertex}, shown in Figure \ref{shrink}. The vertex that is being shrunk can be of any color, as long as it is $2$-valent. The inverse transformation is also allowed and is known as \emph{vertex splitting}. For two graphs $\Gamma$ and $\Gamma'$ related by such a transformation there is a canonical isomorphism $H^1(\Gamma; \K^*) \simeq H^1(\Gamma'; \K^*)$ given by deformation of $\Gamma$ into $\Gamma'$ inside the torus. That isomorphism preserves both the Poisson structure and Goncharov-Kenyon Hamiltonians. Furthermore, the graphs $\Gamma$ and $\Gamma'$ have the same dual quiver $\Q$, and at the quiver level the map $H^1(\Gamma; \K^*) \to H^1(\Gamma'; \K^*)$ is the identity transformation. In other words, the following diagram commutes:

\begin{equation}
\begin{tikzcd}
H^1(\Gamma; \K^*) \arrow[rr] \arrow[dr, "\delta", swap]
& & H^1(\Gamma'; \K^*) \arrow[dl, "\delta'"] \\
 &Y_\Q 
&  
\end{tikzcd}
\end{equation}

 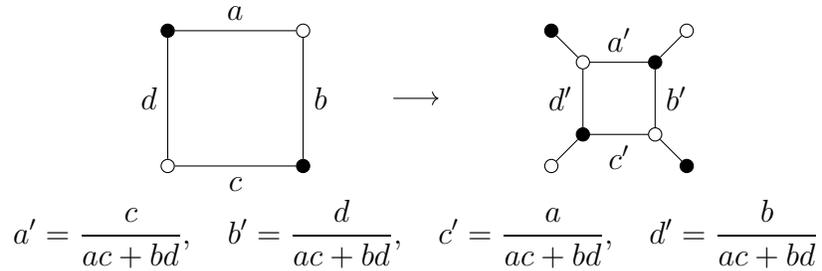
\begin{figure}[t]
 \centering
\begin{tikzpicture}[scale = 0.6]
\node [draw,circle,color=black, fill=white,inner sep=0pt,minimum size=5pt] (A) at (-8.5,0) {};
\node [draw,circle,color=black, fill=black,inner sep=0pt,minimum size=5pt] (B) at (-5.5,0) {};
\node [draw,circle,color=black, fill=white,inner sep=0pt,minimum size=5pt] (C) at (-5.5,3) {};
\node [draw,circle,color=black, fill=black,inner sep=0pt,minimum size=5pt] (D) at (-8.5,3) {};
\draw [thin] (A) -- (B) node[midway, below] {$c$}; \draw (B) -- (C) node[midway, right] {$b$}; \draw (C) -- (D) node[midway, above] {$a$}; \draw (D) -- (A) node[midway, left] {$d$};;
\node [draw,circle,color=black, fill=white,inner sep=0pt,minimum size=5pt] (A) at (0,0) {};
\node [draw,circle,color=black, fill=black,inner sep=0pt,minimum size=5pt] (B) at (3,0) {};
\node [draw,circle,color=black, fill=white,inner sep=0pt,minimum size=5pt] (C) at (3,3) {};
\node [draw,circle,color=black, fill=black,inner sep=0pt,minimum size=5pt] (D) at (0,3) {};
\node [draw,circle,color=black, fill=black,inner sep=0pt,minimum size=5pt] (A1) at (0.7,0.7) {};
\node [draw,circle,color=black, fill=white,inner sep=0pt,minimum size=5pt] (B1) at (2.3,0.7) {};
\node [draw,circle,color=black, fill=black,inner sep=0pt,minimum size=5pt] (C1) at (2.3,2.3) {};
\node [draw,circle,color=black, fill=white,inner sep=0pt,minimum size=5pt] (D1) at (0.7,2.3) {};
\draw [thin] (A1) -- (B1) node[midway, below] {$c'$}; \draw (B1) -- (C1) node[midway, right] {$b'$}; \draw (C1) -- (D1) node[midway, above] {$a'$}; \draw (D1) -- (A1) node[midway, left] {$d'$};;
\draw (A) -- (A1);
\draw (B) -- (B1);
\draw (C) -- (C1);
\draw (D) -- (D1);
\draw [-> ](-3.5,1.5) -- (-2.5, 1.5);
\node () at (-3, -1.5) {$\displaystyle a' = \frac{c}{ac+bd}, \quad b' = \frac{d}{ac + bd}, \quad c' = \frac{a}{ac + bd}, \quad d' = \frac{b}{ac + bd}$};
\end{tikzpicture}
\caption{An urban renewal. Edges with no weight have weight $1$. The corner vertices remain connected to the rest of the graph.}\label{urban}
\end{figure}

\paragraph{Urban renewal.} The second move type is the \emph{urban renewal} shown in Figure~\ref{urban} (there is also a simplified version of the urban renewal known as the \emph{spider move}; up to vertex shrinking/splitting, the urban renewal and spider move are the same).  It is the dimer version of the square move for networks (Postnikov's move 3, see Figure \ref{moves}). In this case, an isomorphism $H^1(\Gamma; \K^*) \to H^1(\Gamma'; \K^*)$ is provided by the transformation of edge weights displayed in the figure. It is  Poisson and preserves the Hamiltonians. The inverse of the urban renewal is not considered since that move is an involution, up to shrinking of $2$-valent vertices.\par
At the cluster level, an urban renewal is a mutation at the vertex corresponding to the square face where the move is being applied. In other words, we have the following commutative diagram:

\begin{equation}
\begin{tikzcd}
H^1(\Gamma; \K^*) \arrow[rr, "\substack{\text{urban renewal} \\ \text{of $i$'th face}}"] \arrow[d, "\delta"] &
& H^1(\Gamma'; \K^*) \arrow[d, "\delta'"] \\
Y_\Q \arrow[rr, "\mu_i^Y" ]
&&  Y_{\Q'}.
\end{tikzcd}
\end{equation}

Conversely, any mutation of $\Q$ at a $4$-valent vertex gives rise to an urban renewal.

\paragraph{Discrete dimer integrable systems.}
Now assume we have sequence of toric bipartite graphs $\Gamma, \dots, \tilde \Gamma$ where each graph is obtained from the previous one by a move. Suppose also that we have an isomorphism of toric bipartite graphs $\psi \colon \tilde \Gamma \to \Gamma$ (i.e. a graph isomorphism preserving vertex colors which is induced by an orientation-preserving homeomorphism $T^2 \to T^2$). Then the composition $H^1(\Gamma; \K^*) \to \dots \to H^1(\tilde \Gamma; \K^*)$ of maps induced by moves, followed by the map $H^1(\tilde \Gamma; \K^*) \to  H^1( \Gamma; \K^*)$ induced by the isomorphism $\psi$, is a birational Poisson automorphism of $H^1( \Gamma; \K^*)$ onto itself which preserves Goncharov-Kenyon Hamiltonians.  
For minimal graphs that map is integrable (as any Poisson automorphism of an integrable system) and is called a \emph{discrete dimer integrable system}. Thus, a discrete dimer integrable system is a sequence of moves of a toric bipartite graph which restores the initial graph, up to an orientation-preserving homeomorphism of the torus onto itself. At the cluster level, it corresponds to a cluster map $Y_\Q \to Y_\Q$, where $\Q$ is the dual quiver of the graph~$\Gamma$. Thus we can view the so-obtained discrete integrable systems as {cluster integrable systems}.

\paragraph{Dimer integrable systems and Newton polygons.}
Note that continuous Goncharov-Kenyon systems associated with graphs related by a sequence of moves are isomorphic.  One can glue such isomorphic systems along moves and hence view them as one system. Regarded in this way, Goncharov-Kenyon systems are classified by their Newton polygons. In other words, two minimal bipartite graphs have the same Newton polygon (up to an integral affine transformation) if and only if they are related by a sequence of moves.

\section{$T$-systems on bipartite quivers}\label{sec:GP}

For this section we picked one of many approaches to generating examples of discrete integrable systems associated with cluster maps. It is due to Galashin and Pylyavskyy \cite{GalPylAJM, GalPylMZ, GalPylDM}. Other constructions can be found, e.g.  in \cite{glick2016meshes, FM}. Rather  than focusing on Arnold-Liouville integrability, \cite{GalPylAJM, GalPylMZ, GalPylDM} explore different flavors of integrability of discrete dynamical systems, namely, periodicity, linearizability,  and algebraic entropy.

\paragraph{\bf $T$-systems on bipartite quivers.} 
As in Section \ref{sec:qmrs}, consider a quiver $\Q$ with vertex set $1, \dots, n$. We say that a quiver is \emph{bipartite} if its vertices are colored black and white so that there are no arrows between vertices of the same color. Following \cite{GalPylAJM}, we call a bipartite quiver $\Q$ {\em recurrent} if both the composition of mutations at all black vertices and composition of mutations at all white vertices result in a quiver $\Q^{op}$ that is obtained from $\Q$ by reversing orientation of all edges, see Figure \ref{FigRec}.   The  composition of the corresponding $X$-mutations is called the {\em T-system} associated with $\Q$. More precisely, the steps of the $T$-system are as follows: $X$-mutation at all white vertices, $X$-mutation at all black vertices, $X$-mutation at all white vertices, etc. Denote by $T_i(t)$ the value of the $x$-variable associated with the vertex $i$ after $t$ steps. Since $T_i(t)$ only changes once in two steps, it is convenient to  assume that $T_i(t)$ is only defined for even times $t$ if $i$ is white and odd times $t$ if $i$ is black. Then

\begin{figure}[t]
\centering
\begin{tikzpicture}[]
\node () at (0,0)
{
\begin{tikzpicture}[scale = 1.3]
\node () at (-1,1) {$\Q$};
\node [draw,circle,color=black, fill=black,inner sep=0pt,minimum size=5pt] (A) at (0,0) {};
\node  [draw,circle,color=black, fill=black,inner sep=0pt,minimum size=5pt] (B) at (0,1) {};
\node [draw,circle,color=black, fill=white,inner sep=0pt,minimum size=5pt]  (C) at (1,1) {};
\node(D) [draw,circle,color=black, fill=white,inner sep=0pt,minimum size=5pt] (D)  at (1,0) {};
\draw [->] (B) -- (C);
\draw [->] (C) -- (A);
\draw  [->] (D) -- (B);
\draw [->] (A) -- (D);
\end{tikzpicture}
};
\node () at (3,0) {$\longrightarrow$};
\node () at (6,0)
{
\begin{tikzpicture}[scale = 1.3]
\node () at (2,1) {$\Q^{op}$};
\node [draw,circle,color=black, fill=black,inner sep=0pt,minimum size=5pt] (A) at (0,0) {};
\node  [draw,circle,color=black, fill=black,inner sep=0pt,minimum size=5pt] (B) at (0,1) {};
\node [draw,circle,color=black, fill=white,inner sep=0pt,minimum size=5pt]  (C) at (1,1) {};
\node(D) [draw,circle,color=black, fill=white,inner sep=0pt,minimum size=5pt] (D)  at (1,0) {};
\draw [->] (C) -- (B);
\draw [->] (A) -- (C);
\draw  [->] (B) -- (D);
\draw [->] (D) -- (A);
\end{tikzpicture}
};
\end{tikzpicture}
\caption{Mutation of a recurrent bipartite quiver at all white vertices.}\label{FigRec}
\end{figure}
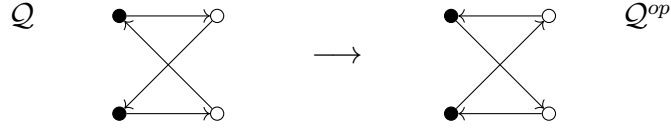
\begin{equation}
\label{Tsys}
T_i(t+1)T_i(t-1)= \prod_{j=1}^n T_j(t)^{\#j \to i} \,+\,  \prod_{j=1}^n T_j(t)^{\#i \to j}
\end{equation}
for all $i$ and $t$ such that the terms are well-defined (i.e. $t$ is odd if $i$ is white and even if $i$ is black). As before, $\#i \to j$ stands for the number of arrows from $i$ to~$j$.

\paragraph{Bipartite bigraphs and subadditive labelings.} To each bipartite quiver $\Q$ one can associate a {\em bipartite bigraph} $(\Gamma(\Q),\Delta(Q))$ defined as follows. Both $\Gamma(\Q)$ and $\Delta(\Q)$ are undirected bipartite graphs with the same vertex set as $\Q$. Edges of $\Gamma(\Q)$ are white-to-black edges of $\Q$, while edges of $\Delta(\Q)$ are black-to-white edges of $\Q$. 
In terms of the bigraph, the $T$-system \eqref{Tsys} can be rewritten as
\begin{equation}
\label{Tsys_bigraph}
T_i(t+1)T_i(t-1)= \prod_{j=1}^n T_j(t)^{\gamma_{ij}} \,+\,  \prod_{j=1}^n T_j(t)^{\delta_{ij}}
\end{equation}
where $\gamma_{ij}$ (respectively, $\delta_{ij}$) is the $(i,j)$ entry of the adjacency matrix of $\Gamma$ (respectively, of $\Delta$), i.e. the number of edges joining vertices $i$ and $j$.

To investigate integrability properties of the system \eqref{Tsys_bigraph}, Galashin and Pylyavskyy studied {\em labelings} of a bipartite bigraph, that is, functions on  its vertex set (i.e. the common vertex set of the graphs $\Gamma$ and $\Delta$) with nonnegative real values. 
A labeling $\nu$ is called 
\begin{enumerate}
\item {\em strictly subadditive} if for any vertex $i$ one has $ 2\nu(i) > \sum_{(i,j)\in \Gamma} \nu(j)$ and $2\nu(i) > \sum_{(i,j)\in \Delta} \nu(j)$ (here and below sums $\sum_{(i,j)\in \Gamma} \nu(j)$, $\sum_{(i,j)\in \Delta} \nu(j)$ should be understood as $\sum_{j=1}^n \gamma_{ij} \nu(j)$ and $\sum_{j=1}^n \delta_{ij} \nu(j)$ respectively);
\item {\em subadditive} if one of the following holds: \begin{itemize} \item either for any vertex $i$, one has  $ 2\nu(i) \geq \sum_{(i,j)\in \Gamma} \nu(j)$ and $2\nu(i) > \sum_{(i,j)\in \Delta} \nu(j)$;
\item or for any vertex $i$ one has  $ 2\nu(i) > \sum_{(i,j)\in \Gamma} \nu(j)$ and $2\nu(i) \geq \sum_{(i,j)\in \Delta} \nu(j)$;
\end{itemize}
\item {\em weakly subadditive} if for any vertex $i$ one has $ 2\nu(i) \geq \sum_{(i,j)\in \Gamma} \nu(j)$ and $2\nu(i) \geq \sum_{(i,j)\in \Delta} \nu(j)$. 
\end{enumerate}

\paragraph{Integrability of $T$-systems.} The most restrictive version of integrability for a discrete dynamical system is periodicity. This may not be so interesting from a dynamical point of view but has deep and important implications in the theory of cluster algebras. The following result is a complete classification of recurrent bipartite quivers that give rise to periodic $T$-systems:

\begin{theorem}\label{Z-period} {\bf{\cite{GalPylAJM}}} The following are equivalent for a recurrent bipartite quiver~$\Q$:
\begin{enumerate}
\item The associated $T$-system \eqref{Tsys_bigraph} is periodic;
\item $\Q$ admits a strictly subadditive labeling;
\item $(\Gamma(\Q),\Delta(\Q))$ is an admissible ADE bigraph, i.e. all connected components of both $\Gamma(\Q)$ and $\Delta(\Q)$ are finite ADE Dynkin diagrams.
\end{enumerate}
\end{theorem}

A weaker integrability-type property is {\em linearizability}: there exists a finite order linear recursion with coefficients given by rational functions in terms of initial variables $x_i$ (or, equivalently, $T_i(0)$ and $T_i(1)$) satisfied by $T_i(t)$ for all $i$.
The discrete open Toda system is an example of a linearizable discrete system. T-systems with this property were termed {\em Zamolodchikov integrable} in \cite{GalPylAJM}.
\begin{theorem}\label{Z-integrable} {\bf \cite{GalPylMZ}}
Let $\Q$ be a recurrent bipartite quiver.\begin{enumerate}
\item Suppose that the $T$-system associated with $\Q$ is Zamolodchikov integrable. Then $\Q$  admits a subadditive labeling.
\item  The quiver $\Q$ admits a subadditive labeling but not a strictly subadditive one if and only if all connected components of one of the graphs $\Gamma(\Q)$, $\Delta(\Q)$ are finite ADE Dynkin diagrams while all components of  the other are affine ADE Dynkin diagrams.
\end{enumerate}

\end{theorem}
\begin{remark}
Here by affine ADE Dynkin diagrams we mean $\tilde A_n$, $\tilde D_n$, $\tilde E_6$, $\tilde E_7$, and $\tilde E_8$. Twisted affine diagrams are not allowed.
\end{remark}
Finally, the weakest (as compared to periodicity and Zamolodchikov integrability) integrability-type condition for discrete systems is \emph{zero algebraic entropy}. Algebraic entropy measures how fast the degrees of rational functions obtained by iteration of the map grow. More precisely, let $\phi: \mathbb{C}^N \to \mathbb{C}^N $ be a rational map, and let $d_n$ be the degree of $n$th iterate of $\phi$ ({to define the degree $\mathrm{deg}\,\psi$ of a rational map $\psi \colon  \mathbb{C}^N \to \mathbb{C}^N$ we write the components of $\psi$ as $\psi_i / \psi_0$ where $\psi_0$ is the least degree common denominator, and let   $\mathrm{deg}\,\psi:= \max_{0 \leq i \leq N } \mathrm{deg}\, \psi_i$}). Then the algebraic entropy of $\phi$ is defined as
\[
\eps = \lim_{n\to\infty}\frac{\log d_n}{n}\ 
\]
(we have an obvious estimate $d_{m+n} \leq d_md_n$, so this limit exists by Fekete's subadditive lemma); see~\cite{bellon1999algebraic}. 

Zero algebraic entropy is a strong indication of integrability in the Arnold-Liouville sense. The results of Galashin and Pylyavskyy on algebraic entropy of $T$-systems  \eqref{Tsys_bigraph} can be summarized in the following
\begin{theorem}
\label{T-entropy} { \bf \cite{GalPylDM,GalPylMZ} }Let $\Q$ be a recurrent bipartite quiver.\begin{enumerate}
\item Suppose that the $T$-system associated with $\Q$ has zero algebraic entropy. Then $\Q$  admits a weakly subadditive labeling.
\item  The quiver $\Q$ admits a weakly subadditive labeling but not a subadditive one if and only if all connected components of $\Gamma(\Q)$ and $\Delta(\Q)$ are affine ADE Dynkin diagrams.
\end{enumerate}
\end{theorem}
Furthermore, \cite{GalPylDM} gives a complete classification of  recurrent bipartite quivers that admit a weakly subadditive labeling and provides detailed conjectures on the converse to the first claim of Theorem \ref{T-entropy} and on 
Arnold-Liouville integrability of $Y$-type cluster maps for quivers admitting weakly subadditive labelings. These conjectures were verified in several important instances. We also refer to~\cite{fordy2014discrete} for results on algebraic entropy of cluster maps beyond the bipartite setting.

\section{Related results}\label{sec:further}

In this section we provide a (by no means exhaustive) list of  results related to constructions discussed above, as well as some particular examples of cluster integrable systems that can be described in terms of those constructions.

\paragraph{Cluster algebras and the KP equation.} One of the first connections between cluster algebras and integrable systems was discovered by Kodama and Williams who related perfect networks\footnote{See Section \ref{xysect}.} (or, more precisely, they undirected counterparts,  \emph{plabic graphs}) to the KP equation \cite{kodama2011kp, kodama2014kp}. In that setting, perfect networks arise as \emph{soliton graphs} encoding relative positions of KP solitons. This relation can also be stated in the cluster language, with cluster variables corresponding to dominant exponentials of soliton solutions.

\paragraph{Cluster algebras and discrete Painlev\'e equations.}
The cluster algebra language can also be used to study nonautonomous analogs of discrete integrable systems, specifically, $q$-Painlev\'e equations \cite{Okubo, hone2014discrete}. For more recent results in this direction, in particular regarding the connection with Goncharov-Kenyon systems, see \cite{bershtein2018cluster, okubo2022generalized}.

\paragraph{Dimers, Poisson-Lie groups, and networks.}
Fock and Marshakov \cite{FM} showed that Goncharov-Kenyon systems (described in Section \ref{sec:GK} above) can be defined via restriction of central functions to Poisson submanifolds of an affine Lie group in type A, and also described all Poisson submanifolds that arise in this fashion in terms of words in the corresponding Weyl group. Furthermore, as shown in~\cite{izosimov2022dimers}, the embedding of a Goncharov-Kenyon system into a Poisson-Lie group can be done via turning a bipartite graph into a weighted directed network and taking the boundary measurement matrix. Furthermore, the class of dimer integrable systems coincides with the ones that can be constructed from a weighted directed network on a cylinder, as described in Section \ref{xysect}.

\paragraph{Quantum cluster integrable systems.} 
Both continuous and discrete dimer integrable systems have quantum counterparts \cite{GK}. It seems to be not known whether those quantum systems can be described in the network language. Quantum networks are studied e.g. in \cite{schrader2017continuous, chekhov2020darboux}.

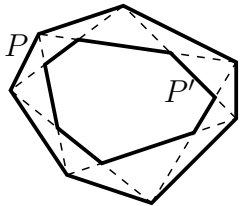
\begin{figure}[t]
\centering
\begin{tikzpicture}[scale = 0.75]
\coordinate (VK7) at (0,0);
\coordinate (VK6) at (1.5,-0.5);
\coordinate (VK5) at (3,1);
\coordinate (VK4) at (3,2);
\coordinate (VK3) at (1,3);
\coordinate (VK2) at (-0.5,2.5);
\coordinate (VK1) at (-1,1.5);

\draw  [line width=0.5mm]  (VK7) -- (VK6) -- (VK5) -- (VK4) -- (VK3) -- (VK2) -- (VK1) -- cycle;
\draw [dashed, line width=0.2mm, name path=AC] (VK7) -- (VK5);
\draw [dashed,line width=0.2mm, name path=BD] (VK6) -- (VK4);
\draw [dashed,line width=0.2mm, name path=CE] (VK5) -- (VK3);
\draw [dashed,line width=0.2mm, name path=DF] (VK4) -- (VK2);
\draw [dashed,line width=0.2mm, name path=EG] (VK3) -- (VK1);
\draw [dashed,line width=0.2mm, name path=FA] (VK2) -- (VK7);
\draw [dashed,line width=0.2mm, name path=GB] (VK1) -- (VK6);

\path [name intersections={of=AC and BD,by=Bp}];
\path [name intersections={of=BD and CE,by=Cp}];
\path [name intersections={of=CE and DF,by=Dp}];
\path [name intersections={of=DF and EG,by=Ep}];
\path [name intersections={of=EG and FA,by=Fp}];
\path [name intersections={of=FA and GB,by=Gp}];
\path [name intersections={of=GB and AC,by=Ap}];

\draw  [line width=0.5mm]  (Ap) -- (Bp) -- (Cp) -- (Dp) -- (Ep) -- (Fp) -- (Gp) -- cycle;

\node at (-0.9,2.3) () {$P$};
\node at (2,1.5) () {$P'$};

\end{tikzpicture}
\caption{The pentagram map.}\label{Figpent}
\end{figure}
\paragraph{Pentagram maps.} One of the most famous examples of cluster integrable systems is the \emph{pentagram map}, introduced by Schwartz in \cite{Sch}. It is a discrete dynamical on the space of projective equivalence classes of planar polygons whose definition is illustrated in Figure \ref{Figpent}: the image of the polygon $P$ under the pentagram map is the polygon $P'$ whose vertices are the intersection points of consecutive shortest diagonals of~$P$ (i.e.,  diagonals connecting second-nearest vertices). 
Integrability of the pentagram map was established in \cite{ovsienko2010pentagram, ovsienko2013liouville, soloviev2013integrability}. A connection with cluster algebras was found in~\cite{glick2011pentagram}. The paper \cite{GSTV} gives a description of the pentagram map in terms of networks (see Figure \ref{FigPSIDO} above) which also extends to certain higher-dimensional generalizations. The dimer model description can be found in \cite{FM} and, in a more geometric setting, in \cite{affolter2019vector}. Further higher-dimensional generalizations of the pentagram map are described in \cite{khesin2013integrability, khesin2015geometry}. Those generalizations also admit a Poisson-Lie interpretation \cite{izosimov2022pentagram}, but a cluster description of such higher pentagram maps remains an open problem.

\paragraph{Classification of discrete dimer integrable systems.}  George and Inchiostro \cite{george2019cluster} classified discrete dimer integrable systems for a given bipartite toric graph (i.e. sequences of local moves restoring the initial graph, up to an orientation-preserving homeomorphism of the torus), modulo those that are identical at the cluster level (i.e. induce identical cluster maps of $Y_\Q$ onto itself). It turns out that if the Newton polygon is an $n$-gon with at least one interior point, then discrete dimer integrable systems on a given graph form an Abelian group of rank $n-3$, confirming an earlier conjecture of Fock and Marshakov \cite{FM}. This description was further refined in \cite{george2022discrete} to allow for certain non-local moves, known as $R$-matrix transformations.

\paragraph{Noncommutative networks.}
Arthamonov, Ovenhouse, and Shapiro extended the network construction described in Section \ref{xysect} above to the case of non-commutative weights \cite{arthamonov2020noncommutative}. In that setting one endows the space of edge weights with a double  bracket in the sense of \cite{van2008double} and shows that it induces an $R$-matrix type double bracket on boundary measurement matrices. In \cite{ovenhouse2020non} such non-commutative networks are applied to study Grassmannian analogues of the pentagram map defined in \cite{felipe2019pentagram}.

\addcontentsline{toc}{section}{References}
\bibliographystyle{plain}
\bibliography{main.bib}

\end{document}